\RequirePackage{lineno}
\documentclass{nature}

\usepackage{amssymb}
\usepackage{graphicx}
\usepackage{epstopdf}
\usepackage[FIGTOPCAP]{subfigure}
\usepackage{color}
\usepackage[normalem]{ulem}
\usepackage{soul}
\usepackage{overpic}
\usepackage[position=above]{caption}
\usepackage{amsmath}
\usepackage{mathrsfs}
\usepackage{booktabs}

\def\nat{Nature\ }
\def\aap{Astron.\ Astrophys.\ }
\def\apj{Astrophys.\ J.\ }
\def\apjl{Astrophys.\ J.\ Lett.\ }
\def\apjs{Astrophys.\ J.\ Supp.\ }

\def\mnras{Mon.\ Not.\ Roy.\ Astron.\ Soc.\ }
\def\physrep{Phys.\ Rept.\ }
\def\prd{Phys.\ Rev.\ D\ }

\def\araa{Annu.\ Rev.\ Astron.\ Astrophys.\ }

\def\cjaa{Chin.\ J.\ Astron.\ Astrophys.\ }

\def\pasp{Publ.\ Astron.\ Soc.\ Pac.\ }
\def\ssr{Space Sci. Rev.}   

\usepackage{floatrow}
\DeclareFloatFont{tiny}{\scriptsize}
\title{{An optical-ultraviolet flare with absolute AB magnitude of -39.4 detected in GRB 220101A}}

\author{Zhi-Ping Jin$^{1,2,3}$, Hao Zhou$^{1,2,3}$, Yun Wang$^{1,3}$,  Jin-Jun Geng$^{1}$,
Stefano Covino$^{4}$, Xue-Feng Wu$^{1,3}$, Xiang Li$^{1}$, Yi-Zhong~Fan$^{1,2,3}$,  Da-Ming Wei$^{1,2,3}$, and Jian-Yan Wei$^{5}$.}

\begin{document}

\maketitle

\begin{affiliations}
\small
\item{Purple Mountain Observatory, Chinese Academy of Sciences, Nanjing 210023, China}
\item{Key Laboratory of Dark Matter and Space Astronomy of Chinese Academy of Sciences, Nanjing 210023, China}
\item{School of Astronomy and Space Science, University of Science and Technology of China, Hefei 230026, China}
\item{INAF/Brera Astronomical Observatory, via Bianchi 46, I-23807 Merate (LC), Italy}
\item{National Astronomical Observatories, Chinese Academy of Sciences, Beijing 100049, China}
\end{affiliations}


\begin{abstract}
Hyper-luminous optical/ultraviolet flares have been detected in Gamma-ray Bursts and the {luminosity} record was held by naked eye event GRB 080319B. Such flares are widely attributed to internal shock or external reverse shock radiation. 
With a new dedicated method developed to derive reliable photometry from saturated sources of {\it Swift}/UVOT, here we carry out time-resolved analysis of the initial White band $150~{\rm s}$ exposure of GRB 220101A, a burst at the redshift of 4.618, and report a rapidly-evolving optical/ultraviolet flare with a high 
absolute AB magnitude of $-39.4 \pm0.2$. At variance with GRB 080319B, the temporal behavior of this new flare does not trace the gamma-ray activities. Rather than either internal shocks or reverse shock, this extremely energetic
optical/ultraviolet {flare} is most likely from the refreshed shocks induced by the catching-up of the late-ejected extremely-energetic material with the earlier-launched decelerating outflow.  {This finding reveals the diverse origins of the extremely energetic ultraviolet/optical flares and demonstrates the necessarity of the high time-resolution observation at early times.} 

\end{abstract}

Gamma-ray bursts are widely believed to originate from the internal energy dissipation of a highly relativistic and narrowly collimated outflow that was launched by a nascent stellar mass black hole or magnetized neutron star \cite{1995ARA&A..33..415F,2002ARA&A..40..137M,2004RvMP...76.1143P,2015PhR...561....1K}. Shortly after the onset of prompt emission of GRBs, there could come very bright optical/ultraviolet flares arising from either the internal shocks in specific conditions or the external reverse shock radiation\cite{Meszaros1997}.  An apparent  $\sim 9$th mag optical radiation peak was detected in GRB 990123 at a redshift of $z=1.62$\cite{Akerlof1999,1999Sci...283.2069C}. Its rapid rise and the  quick decline are consistent with the reverse shock radiation model\cite{Sari1999, Meszaros1999}, and the late more-detailed afterglow modeling revealed that the reverse shock region should be significantly more magnetized than the forward shock region\cite{Fan2002, Zhang2003}. A long-holding {luminosity} record was set by GRB 080319B, a burst at a redshift of $z=0.937$.  Its peak visual magnitude reaches
5.3, which is so bright that an observer in a dark location could have seen it with the naked eyes\cite{Racusin2008}. The correlated temporal behaviors of the prompt gamma-ray emission and the optical radiation are in favor of the internal shock process\cite{Fan2009,Li2008}. However, in the past decade,  no similar or even comparable events have been reported.  
One reason is that the intrinsic duration of the ultra-bright optical/ultraviolet emission (${\cal T}_{\rm p,uvo}$) of GRB 080319B lasted just about 20 seconds,
and the detection of such transients needs an efficient and rapid alert system and telescopes able to rapidly slew to the target, which are quite challenging. Indeed,  the detection of the extraordinarily-bright optical emission of GRB 080319B was a coincidence. About 30 minutes earlier, GRB 080319A took place in the direction that was only 10 degrees away from GRB 080319B.  Several wide-field small-size telescopes were hunting for the optical afterglow of GRB 080319A and fortunately caught the optical counterpart of GRB 080319B instantly. 

Interestingly and likely also ``surprisingly", 
GRB 080319B, if it had taken place at a high redshift $z\sim 5$, it might have been detectable by {\it Swift} ultraviolet and optical telescope (UVOT) with the White and possibly also V filters. First of all, at such a high redshift, the peak optical emission would be extended to an observation time of $\sim (1+z){\cal T}_{\rm p,uvo}\sim 100$ s, while the typical slew time of the UVOT to the target is about  80 s\cite{Roming2017}. 
Therefore,  during $\sim 100$ s a prompt slew of the UVOT to the object is plausible. In addition,  the luminosity distance $D_{\rm L}\sim 1.5\times 10^{29}$ cm for a redshift of $z\sim 5$ will make the source dimmer by a factor of $\sim D^{2}_{{\rm L}(z=5)}/D^{2}_{{\rm L}(z=0.937)} \sim 57$ and the Lyman alpha absorption will further decrease the observed flux considerably (i.e., the correction  is about 2 magnitudes in V band and reaches {$\sim 5$ magnitudes for the White filter} because of the bluer range that also alleviates the risk of saturation).
Finally, for saturated sources with associated readout streaks in the image,
the proper correction can yield reliable photometric measurements\cite{Page2013} about 2 mag brighter than can be achieved using the standard aperture photometry with a point-source correction for coincidence loss\cite{Poole2008,Kuin2008}.
Anyhow, for $z>5$, the observation is challenging because the Lyman alpha absorption appearing at $\approx 7296$ \AA$ [(1+z)/6]^{-1}$ 
is already beyond the red edge of the wavelength range of UVOT/White filter.  As demonstrated below,  GRB 220101A represents a remarkable and also the first example for the UVOT detection of an extraordinarily-luminous optical/ultraviolet flare, which is actually the most energetic one detected so far.

GRB 220101A was discovered simultaneously by {\it Swift} Burst Alert Telescope (BAT)\cite{2022GCN.31347....1T},  the Fermi satellite (both the Gamma-ray Burst Monitor and the Large Area Telescope)\cite{2022GCN.31360....1L} and 
the AGILE satellite\cite{2022GCN.31354....1U}.  Before the so-called finding chart exposure ranging from 90 to 240 seconds with the White filter\cite{2022GCN.31347....1T}, UVOT observed the target in V band for 9 seconds  starting 71 seconds after the BAT trigger.  
The estimated average magnitude in the White band for an exposure of $\sim 150$ s is $\sim 14.7$th Vega mag\cite{2022GCN.31347....1T,2022GCN.31351....1K} (i.e. 15.5th AB mag). The field of GRB 220101A  was observed by the Xinglong-2.16m telescope equipped with the BFOSC camera. The redshift was measured to be $z=4.618$\cite{2022GCN.31353....1F,2022GCN.31359....1F} and in the spectrum a broad absorption feature, which results from the Lyman alpha absorption, is evident centered at $\sim 6820$ \AA. 
The corresponding isotropic equivalent gamma-ray energy is $\sim 4\times 10^{54}$ erg and the peak luminosity is $\sim 9\times 10^{53}~{\rm erg~s^{-1}}$, both are in the rank of the most energetic ones among known GRBs\cite{2022GCN.31365....1A,2022GCN.31433....1T}. After the redshift correction, the observed optical photons were intrinsically in the ultraviolet bands. Therefore, all the emission detected by {\it Swift} suffered from serious absorption (in the observer's frame, the V band absorption is about 2 mag stronger than that in the I-band, as found in Ref.\cite{2022GCN.31353....1F}) and thus the intrinsic emission would be much brighter. This is in particular the case for the White filter because of its large effective area in the blue part (i.e., U, UVW1, UVM2 and UVW2) and the Lyman alpha/intergalactic medium (IGM) absorption would be very strong. Therefore, in this work we analyze the White filter data to determine its intrinsic brightness. Here we concentrate on the first $\sim 150$ s exposure with White filter though it seems to be unsaturated, the reason is that the very early optical/ultraviolet emission of GRBs usually show strong variability and such information can only be revealed by  high time resolution data analysis.  Fortunately, the first epoch UVOT/White observation was in the event mode  (i.e. photon counting mode)  and can be efficiently divided into short bins according to the signal-to-noise ratio (SNR).  Our time-resolved analysis did reveal that the measurements in the time range of $\sim 106-150$ s after the BAT trigger suffered from strong saturation, as shown in Fig. 1, Supplementary Table 1 and Extended data Fig. 1.  The absence of clear signal of the read-out streaks in the raw data,  indicating a moderate saturation, however hampers a correction following procedures  discussed in the literature\cite{Page2013}. Therefore, in this work we propose a new method  to correct the saturation effect. The basic idea is that though the pile up at the source site is so serious that can not be reliably corrected, the surrounding but relatively ``separated" pixels are possibly unaffected by saturation and therefore the enhancement of the counts should be correlated with the intrinsic count rate of the source (in other words, the outer part of the point spread function of the source image in the CCD is unchanged). To check whether it is the case, we need some data with known magnitudes as well as the count rates in external annuli. For the unsaturated data with relatively low ring count rate, we simply take UVOT/White measurements of GRB 220101A at {$150-240$} seconds after the burst trigger. For the moderate saturation that is of our great interest, we take the UVOT measurements of GRB 130427A in the time interval of $500-2000$ seconds. 
Though the moderately saturated White band emission of GRB 130427A can not be directly measured, we infer them with the UVOT emission in other bands since the spectrum can be well fitted by a single power-law, see Extended data Fig. 2. 
With these two sets of data, we do find a tight correlation between photon count rate in $5''$ aperture ($\dot{N}_{\rm aper}$, directly measured if unsaturated, or inferred from the ``intrinsic" count rate $\dot{N}_{\rm int}$ measured in other ways) and in the $15''-25''$  ring ($\dot{N}_{\rm ring}$, directly measured in UVOT images), which reads $\dot{N}_{\rm aper}=(22.22\pm0.84)\dot{N}_{\rm ring}$ for $\dot{N}_{\rm ring}\leq 80~{\rm s}^{-1}$ (see Extended data Fig. 3). The correlation efficient for such an empirical relation is 0.99, see the Methods for the extensive discussion, including the further independent verification with some bright stars. The other essential correction is on the absorption of the ultraviolet photons at high redshift. In the analysis we correct such a factor with the wide band energy spectrum and further check it with the other two GRBs at rather similar redshifts (see the Extended data Fig. 4 and Supplementary Table 1).

In Fig. 1 we show the lightcurves of the prompt gamma-ray emission and the very early optical emission. The first White exposure with a duration of 150 s was in the event mode. Thanks to the brightness of the flare, the data analysis can be carried out in much shorter time intervals. 
In our approach, a bin size of $4$s is adopted. 
In principle, a narrower bin size is helpful in revealing the peak or structure of the flare, but a reasonably wide bin 
is necessary for a high SNR (comparing to the $5''$ aperture, in the ring there are only $\sim5\%$ signal photons but 16 times background photons). 
The optical/ultraviolet flare lightcurve is relatively smooth and 
there is no evidence for 
tracing the temporal behavior of prompt gamma-rays. 
This is very different from the case of GRB 080319B, where the naked-eye optical flash shows strong variability and  resembles the gamma-ray lightcurve (see the insert of Fig. 1), which will shed valuable light on the physical origin. 
We have also constructed the ``prompt" optical, X-ray and gamma-ray SED. 
In Fig. 2 we just show three representative time intervals of the first UVOT White band exposure, including the very beginning, the peak, and the final shallow decline phase. In the rise and the quick decline phases, the extrapolation of the X-ray spectrum into the optical is (well) below the White band measurements, which again points towards different physical origins of the optical and high energy radiation. While in the $t^{-2.3\pm0.3}$ shallow decline phase, the optical to X-ray emission are consistent with being a single power-law, which may be dominated by the external reverse shock radiation. 
In Fig. 3 we present the absolute AB magnitudes of the very early optical emission of GRB 220101A and the other three remarkable events (i.e., GRB 990123\cite{Akerlof1999}, GRB 050904\cite{Boer2006} and GRB 080319B\cite{Racusin2008})
 characterized by the very bright optical emission. Interestingly, before 2022 no optical/ultraviolet flash as energetic as GRB 080319B had been observed and the second record is dimmer by a factor of $\sim 5$. After the proper saturation, absorption as well as the cosmological corrections, it turns out that GRB 220101A sets a new record. Note that the peak optical emission of GRB 220101A could be even stronger than presented here since our current fluxes are the average of the radiation in each 4s bin.   

 As already mentioned before, for the naked-eye optical flare detected in GRB 080319B, the internal shock model is favored by the similar temporal behaviors of the prompt gamma-ray and optical radiation. While for GRB 990123, the external reverse shock model has been widely accepted to account for the optical flash. The optical/ultraviolet flare detected in GRB 220101A, however, should have a different physical origin for the following facts: (i) In contrast to GRB 080319B, the optical flare lightcurve of GRB 220101A does not trace the variability of the prompt gamma-rays (see Fig. 1), requiring different radiation processes/sites of these two components; (ii) The $t^{-3.4}$-like decline of the optical/ultraviolet flare of GRB 220101A may be due to the reverse shock emission, but the $\sim t^{20}$ increase is much quicker than that of GRB 990123 and hence strongly in tension with the standard external reverse/forward shock emission model\cite{Sari1999,Meszaros1999}.    
Here we present a model for the brightest optical spike of GRB 220101A. 
Looking at the gamma-ray lightcurve, the main burst phase consisting of two giant gamma-ray spikes appears at $\sim 90$ s after the BAT trigger, and the earlier emission was much weaker (i.e., the time-averaged luminosity is $\sim 10^{52}~{\rm erg~s^{-1}}$). As indicated by the bulk Lorentz factor$-$luminosity correlation\cite{Lv2012,Fan2012} of $\Gamma \propto L_\gamma^{0.3}$, the weak/slow GRB outflow component launched at the early times is expected to have a $\Gamma \sim {\rm a~few}\times 10^{2}$ and the surrounding interstellar medium further decelerates the outflow to a Lorentz factor of $\Gamma_{\rm W}$, while the outflow component yielding the most luminous part of GRB 220101A likely has a Lorentz factor of $\Gamma_{\rm M}\sim  10^3$. The first giant spike comes from the energy release of the main outflow, either through the internal shocks or the magnetic reconnection within it. Soon the main outflow would catch up with the decelerating weak part at a time of $\sim \Gamma_{\rm W}^{2}\delta t_{\rm WM}/\Gamma_{\rm M}^{2}\sim {\cal O}(10)$ s, which explains the second gamma-ray spike and the delayed onset of the optical/ultraviolet flare, where $\delta t_{\rm WM} \sim 100$ s is the delay of the onset of the main part with respect to that of the weak part (started at $\sim 60$ s before the trigger, see Fig. 1).
The collision of the late/fast material shell(s) with the early/decelerating material will generate strong internal shocks and then produce energetic emissions. Following the treatments presented in Sec. 2.1 of the Ref.\cite{Wei2006}, it is straightforward to show that for the internal shocks taking place at $\sim 
10^{16}~{\rm cm}~(\Gamma_{\rm W}/10^{2})^2 (\delta t_{\rm WM}/10^{2}~{\rm s})$, the typical synchrotron radiation frequency is indeed within the optical/ultraviolet bands. The bulk Lorentz factor of the merged shells can be approximated to be $\bar{\Gamma} \approx \sqrt{[M_{\rm W}\Gamma_{\rm W}+M_{\rm M}\Gamma_{\rm M}]/[M_{\rm W}/\Gamma_{\rm W}+M_{\rm M}/\Gamma_{\rm M}]}$ and the Lorentz factor of the internal shocks can be estimated as $\Gamma_{\rm sh}\approx \Gamma_{\rm M}/\bar{\Gamma}+\bar{\Gamma}/\Gamma_{\rm M}$, where $M_{\rm W}$ and $M_{\rm M}$ are the rest masses of the ejecta powering earlier weak gamma-ray emission and the main outburst, respectively\cite{Piran1999}. Indeed, for GRB 220101A-like burst, we have the outflow luminosity of $L_{\rm m} \sim 10^{53}-10^{54}~{\rm erg~s^{-1}}$, with the fractions of the shock energy given to the magnetic fields (electrons) $\epsilon_{\rm B} \sim 0.1$ ($\epsilon_{\rm e} \sim 0.3$), $\bar{\Gamma} \sim {\rm several\times 100}$ and $\Gamma_{\rm sh}\sim$ a few, it is natural to have an optical/ultraviolet flux\cite{Wei2006} of $\sim 1$ Jy even for a redshift as high as $\sim 5$. In the Methods {and Extended Data Fig.5}, we show numerically that such a process is able to reproduce the data.

Note that the very energetic prompt emission appearing at $\sim T_0+90$ s after the BAT trigger, which partly overlap with the optical/ultraviolet flare, should also effectively cool the electrons accelerated in the collision discussed above. Such a process would produce GeV emission, which is expected to last longer than the overlapping phase of the prompt MeV emission and ultraviolet/optical flare. Indeed, at $t\sim 100-150$ s after the BAT trigger, GeV emission was detected from GRB 220101A\cite{2022ApJ...941...82M}. 

Though the  hyper-luminous very early optical/ultraviolet emission are not common, we suggest that the bursts with prompt emission resembling GRB 220101A (i.e., the much more energetic outbursts appeared at late times and the early emission are well separated from the late ones) are good candidates. 
The problem is how to catch such signals promptly.  Small telescopes with a large field of view should be very helpful and the $I/R$-band observation of these telescopes can catch the monsters in a wide range of redshifts. Anyhow, such observations are limited by the weather, the time (day or night) and the burst site. The space telescopes such as {\it Swift}/UVOT and SVOM/VT\cite{Yu2020} may play an important role in detecting the high redshift events. Since the optical/ultraviolet flare of GRB 220101A was observed by  {\it Swift}/UVOT, below we focus on the upcoming 0.4m SVOM/VT with two channels, including the blue ($400-650$ nm) and the red ($650-1000$ nm) bands. For the shortest exposure time of $1$s, the saturation limit is about 9th magnitude. Given its higher sensitivity in comparison to {\it Swift}/UVOT V filter,  the seriously absorbed ``ultraviolet" emission of GRB 220101A/GRB 080319B-like extra-luminous events, even taking place at the even higher redshift (say, $z\sim 6$), can still be caught by the blue channel of SVOM/VT though the red channel might be saturated (see the Methods and also Extended data Fig. 6). 
Dedicated observation strategies are needed to optimize the potential of the discoveries.


\clearpage

\begin{methods}

{\bf A new method to measure the saturated sources in {\it Swift} UVOT images. }
UVOT is a photon counting detector and typical read-out rate is once every $\sim11$ ms. 
If the source is bright enough ($>10~{\rm counts~ s^{-1}}$), coincidence losses start to be significant and a correction is necessary.
When the incident photon counts rate beyond the read-out rate $\sim 86~{\rm s^{-1}}$,
the source is fully saturated and proper coincidence loss correction is impractical\cite{Poole2008}.
However for extremely saturated sources with read-out streaks, a calibration method has been developed based on the measurement of read-out streak line strength\cite{Page2013}. 
Anyhow, the read-out streak lines are only present in the extremely saturated sources or those with very long time exposure. For the moderate saturation with relatively short exposure, 
it cannot be applied and our main goal  is to provide a new way.
Below we focus on the White band, but our method can be applied to other UVOT filters as well (indeed, as a validation, we also show in the end of this subsection that a rather similar empirical correction function holds for the V band).

The saturated pattern of an UVOT image can be divided into three parts. The first is a point source like structure at the center of saturated pattern, which represents the location of the saturated source. The second part is a dark square structure caused by coincidence loss and the half length of its diagonal line is about 14 arcsec. A more detailed explanation is that UVOT has actually {\it a 256$\times$256 CCD} which records the electron splash pattern produced by the incident photon amplification in a multi-channel plate device and there is a centroid algorithm to calculate positions of incident photons whose accuracy could reach 0.125 pixel\cite{2000MNRAS.319..414F}. As a result, each physical pixel could be subsampled to 8$\times$8 virtual pixels with a resolution of 0.5 arcsec/pixel. The side length of the dark square is about 20 arcsec, that is 40 virtual pixels, corresponding to an area of 5$\times$5 pixels region on real physical CCD which is the affected region of coincidence loss. The third part is the halo ring, which is distinct for saturated sources and some unsaturated sources but with low background. Extended data Fig. 1 shows such a saturated pattern. We attribute the halo rings to  the wing of the Point Spread Function (PSF) of UVOT detector. To test this conjecture, we will examine whether the ``intrinsic" photon counts rates of saturated sources is proportional to photon counts rates of halo rings\cite{2010MNRAS.406.1687B}.

To avoid the influence of the coincidence loss, the best measurement region to get the highest S/N ratio is the area between a circle with a radius of 25 arcsec and a square,  with the same center and with a side length of 20 arcsec. 
However, if {\it Swift} rotated during observations, the dark region of final stacked science image are not necessarily a square due to that the coincidence loss square is aligned to the edge of CCD. Hence, we used an annulus of an inner radius of 15 arcsec and an outer radius of 25 arcsec (i.e., the outer edge of halo rings) to measure photon count rate 
in the ring ($\dot{N}_{\rm ring}$), where the background should be removed and the coincidence loss has been corrected\cite{2000MNRAS.319..414F,Poole2008,Kuin2008,2015MNRAS.449.2514K}. Total observed count rates (both the source and the background) of wing regions of all saturated sources analysed in this paper are less than 105 count/s and for unbinned images, total observed count rate densities are less than 0.021 count/s/pixel, which falls in the typical UVOT background range ($0-0.05$ count/s/pixel). Hence, the coincidence loss correction of the wing region is assumed to be same as that of backgrounds, (i.e., with the area ratio of the standard coincidence loss region for point sources to the wing region, $\frac{5^2\pi}{(25^2-15^2)\pi}=\frac{1}{16}$, convert the total observed count rate in the wing region to an equivalent count rate of standard coincidence loss region, then get the coincidence loss correction factor, which is applied to correct the total count rate of the wing region, with the formula in Poole et al.\cite{Poole2008}). In addition, for cases that background intensities $<$ 0.02 count/s/pixel, the coincidence loss correction formula in Poole et al.\cite{Poole2008}, which is used in Ftools \textit{uvotsource}, can give a reasonable correction\cite{2010MNRAS.406.1687B}. However, different from backgrounds, counts are not spatially uniform in the wing region. By fitting the count distribution in the wing region, we found that the nonuniformity only leads to a difference of $\sim0.8\%$ for the case that total observed count rate is 100 count/s, and the difference will decrease almost linearly with the total observed count rate, which is negligible in our analysis.
The crucial step is to reliably derive the corresponding photon count rate of the saturated source within the standard aperture with a radius of 5 arcsec ($\dot{N}_{\rm aper}$). As mentioned above, if the incident photon counts rate is beyond the CCD readout rate, the source is saturated. Fortunately, the UVOT White band is much wider than other 6 bands (hence, we will call them the narrow bands), which means although a source is saturated in White band, it could be unsaturated in narrow bands. It is therefore plausible to measure the spectrum with other filters of UVOT and then convolve it with the White filter to get the corresponding ``intrinsic" emission. This can be done for the power-law like afterglow spectrum of GRBs and the very early time optical flash of GRB 130427A is a nice sample. The earliest UVOT measurements of this burst were highly saturated and some of them can be analyzed with the readout streak method\cite{Page2013}. Moreover, as shown in Maselli et al.\cite{Maselli2014} and the top panel of Extended data Fig. 2, when the White filter was still saturated, there were 
usable measurements in other bands. 
In the bottom panel of Extended data Fig. 2, we show the ultraviolet/optical SED of GRB 130427A with the UVOT observations. 
Note that these data were re-measured in this work and they are consistent with that reported in the literature\cite{Maselli2014}. We performed the early time photometry of GRB 130424A with \textit{HEASoft} and the results are summarized in Supplementary Table 2. The first exposure in $B$ band and the first 2 exposures in $U$ band were saturated, hence we took the values from Maselli et al\cite{Maselli2014}. 
Light curves of 6 narrow bands were fitted to obtain their magnitudes simultaneous with White band exposures, the results are listed in Supplementary Table 3.
We then carry out the power-law spectral fit to the SED and estimate the White band magnitudes, as summarized in the last column of Supplementary Table 3, which are further used in Supplementary Table 4 to yield the inferred count rate $\dot{N}_{\rm int}$ 
(in another word, the derived $\dot{N}_{\rm aper}$). 
It is worth noting that in epoch 1 there was an optical/ultraviolet flare and hence it is not suitable to evaluate the White band emission with this method. 
Moreover, the White band measurement in the first, second and third epochs were significantly saturated with readout streaks, for which the fluxes were reported before. As show in Extended data Fig. 3, 
in epoch 2  our calculated flux is consistent with that reported in Maselli et al.\cite{Maselli2014}, validating the method proposed in this work. 
Our downloaded image of the epoch 3 mentioned in Maselli et al.\cite{Maselli2014} is distorted and we have hence focused on the subsequent observation data with an exposure of 20 s. Our estimated flux is still well consistent with that reported in Maselli et al.\cite{Maselli2014}, which is expected because these two measurements were almost simultaneous. Anyhow, in the plot the data point reported in Maselli et al.\cite{Maselli2014} is not shown because we can not measure its ring count rate because of distortion.  For epoch 4 to epoch 8, 
there were no readout streaks and the method developed by Page et al.\cite{Page2013} does not work any longer. Our method mentioned above applies to these data and yields reasonable results. 
As for GRB 220101A, shortly after its peak, the ultraviolet/optical flare is not saturated any more. For these observations we can reasonably measure its White band emission. \textit{HEASoft} UVOT Ftools were used to obtain photometry of barely saturated source images of GRB 220101A with a circle aperture with a radius of 5 arcsec. However, a reliable measurement of the ring count rate requires a sufficiently long exposure. Therefore, we just divide the ``tail" part of the flare into two time intervals. We also notice 3 bright stars in the field and then measure them for an independent check.  These five data points are summarized in Supplementary Table 4.
The White band fluxes measured (indirectly and directly, respectively) in the above  events and field stars are used to clarify whether there is a tight correlation between the ring counts and the intrinsic source emission. 
For such a purpose, these three data sets have been fitted with a linear function of  a model of $y=ax$ and a least square cost function was applied, $\chi^2=\sum_{\rm i}\frac{(y_{\rm i}-a x_{\rm i})^2}{y_{\rm err,i}^2+(a x_{\rm err,i})^2}$, where $y_{\rm i}$ and $x_{\rm i}$ represent extracted White-band photon counts rates and halo ring photon counts rates, respectively, and $y_{\rm err,i}$ and $x_{\rm err,i}$ are the corresponding uncertainties. The  Pearson correlation coefficient is 0.99, which reveals a very strong linear correlation, and the $\chi^2/{\rm d.o.f}$ value is $\sim0.90$, which implies a reasonable fit, where ${\rm d.o.f}$ denotes the degree of the freedom.
Hence, we conclude that $\dot{N}_{\rm aper}=22.22\pm 0.84 \dot{N}_{\rm ring}$ can yield a reasonable estimation of ``true" photon counts rates of saturated sources in White band.
Extended data Fig. 3 presents our best fitting result
which confirms our early speculation and suggests that the outer part of the PSFs of such sources are nearly unmodified. 

The ground-based telescopes can well measure the V-band emission of the sources, which can thus provide an economical way to calibrate the saturated V-band observations of {\it Swift}/UVOT. 
Interestingly, GRB 080319B is a nice example. 
For the UVOT V-band observations, in total we have selected 22 time slices in the event files, which were later converted to images with \textit{HEASoft} for measurements. The first 4 exposure duration are 30s, 40s, 50s and 55s, which are same as the time bins in Page et al.\cite{Page2013}. These exposures display readout streaks and have been analyzed with the method of Page et al.\cite{Page2013}, which are shown in the bottom panel of Extended data Fig. 3 (see the light green empty squares). We measured the counts rate in the halo rings, which is defined above, with \textit{HEASoft}, but made coincidence loss correction manually. Another 18 images are unsaturated, the intrinsic emission were directly measured, and they are marked with dark green empty triangles in the bottom panel of Extended data Fig. 3. These measurements are summarized in Supplementary Table 5. In addition, the optical emission of GRB 080319B was well measured by the ground based telescopes \cite{2009ApJ...691..495W}, and the accurately measured V-band emission from RAPTOR-T can be taken as the intrinsic ones (i.e., we have the $\dot{N}_{\rm int}$, in another word, $\dot{N}_{\rm aper}$ defined in this paper). The difference between the V filter of UVOT and that of RAPTOR-T is small and the magnitude difference can be ignored, as demonstrated by the overlapping data points in the left lower corner of the bottom panel of Extended data Fig. 3. Since the very early UVOT/V band observations were in event mode, we can re-bin them into the time intervals the same as that of RAPTOR-T and then get the $\dot{N}_{\rm ring}$. Time bins of our measurements are listed in Supplementary Table 5. Therefore, we apply the linear fit to the data sets and find an empirical function of $\dot{N}_{\rm aper}=20.6\pm 0.4\dot{N}_{\rm ring}$ with a high correlation coefficient of $0.998$. Such a correlation is nicely consistent with that for the UVOT/White band.  It is worth noting that for GRB 220101A, the photons collected in the White band are dominated by those passing the V filter because of the serious absorption in the bluer region. Indeed we find rather similar count rates for the (almost) simultaneous White and V-band measurements (see Supplementary Fig. 1). Therefore, the rather similar correction function for UVOT/V filter strongly 
suggests that our White band analysis of GRB 220101A is robust.  

{\bf {\it Swift} UVOT data analysis.} 
{\it Swift}/UVOT observed GRB 220101A in V, B, U, W1, M2, W2 and White bands for several epochs. 
For data in image mode, we started from the level 2 UVOT products and used standard  aperture photometry, background was measured in a nearby region without sources in stacked images. Reliable detections were only obtained in V and White bands, and the photon count rates were measured in 3 or 5 arcsec apertures, depending on SNR. Coincidences loss correction and aperture correction were applied. For images without detection, upper limits were assuming count rates would have reached the SNR of $S/N=3$. Finally zeropoints including long-term sensitivity correction were used for absolute calibrations. The results are shown in Supplementary Table 1. 

The first white-band exposure under event mode (incident positions and time of every photon are recorded) began at about 90 seconds after the trigger time, which lasted about 150 seconds. Due to the fact that the luminosity of GRB 220101A changed rapidly at early epochs, although the transient seems to be unsaturated on the image for the total 150s exposure, it could be saturated in its peak phase. Hence, we screened the calibrated event data into slices whose exposure time is $\sim$4s to check whether the situation mentioned above had happened. Following the guidance of UVOT data process (https://swift.gsfc.nasa.gov/analysis/UVOT\_swguide\_v2\_2.pdf), event slices were transformed into images and image calibrations (flat field and mod 8 corrections) were applied. Since the transient is bright and isolated on reduced images, standard aperture photometry method was applied.
From 90s to 100s, the transient was brightening rapidly and then became saturated for about 50 seconds. After $\sim$ 150s since the trigger time, it became unsaturated, again. We found that there are halo rings around the transient on barely saturated and saturated images, which we think are the 'wings' of point spread functions, hence, we analyzed the data with our calibration method described before in the Methods. The results are summarized in Supplementary Table 1. 

{\bf {\it Swift}-BAT/XRT and Fermi-GBM data analysis.} 
We processed {\it Swift}-BAT data according to standard procedures, using the software \textit{HEASoft} (ver. 6.29) and calibration database ({\it CALDB}).
The mask weighting file used in extracting the light curve is generated by {\it batgrbproduct} (a complete GRB processing script in \textit{HEASoft}).
We extract event data at time intervals between -60 to 340 seconds related to the trigger time, the energy range is 15-350 keV, and the time bin size is 1 second. 
Our BAT analysis results are plotted with our {\it Swift} UVOT analysis results in Fig. 1. 

We also present a spectral analysis in a broad gamma-ray band (0.3 - 40000 keV) from {\it Swift}-BAT/XRT and Fermi-GBM data.
The files used include the source and background spectrum files, as well as the corresponding response functions.
For BAT file extraction and correction, we used standard procedures as in the manual (https://swift.gsfc.nasa.gov/analysis/swiftbat.pdf).
XRT files were created by online analysis tools provided by {\it Swift} official website \cite{2007A&A...469..379E,evans2009methods}.
The Fermi-GBM data have been processed with {\it GBM Data Tools}. 
There are different statistics used for each dataset ($cstat$ for {\it Swift}-XRT, $\chi^{2}$ for {\it Swift}-BAT and $pgstat$ for Fermi-GBM data).
We use {\it Bilby} in the framework of {\it PyXspec} for model parameter estimation. The results are shown in Fig. 2.

{\bf Intrinsic optical/ultraviolet emission of GRB 220101A.} To estimate a reliable un-absorbed optical/ultraviolet emission, we need an intrinsic spectrum to evaluate the absorption in different observation bands. For such a purpose, in addition to the UVOT V and White band observations, we adopt the $g, r, i, z$-band data from Liverpool telescope measured at $t\sim 0.625$ day after the burst\cite{2022GCN.31357....1P} and the simultaneous XRT data. Such a set of ground-based telescope observation data are chosen because they are almost simultaneous with one UVOT White measurement and at late times the White band emission was not detectable any longer (see Supplementary Table 1 and Supplementary Fig. 1). 
The SED from $i$ to $g$ declines very rapidly, requires a spectral index $\beta\sim 8$ (see Extended data Fig. 4). Similar rapid declines, due to the serious Lyman forest absorption, have been observed in GRB 000131\cite{Andersen2000} and 100219A\cite{Thone2013} at redshifts of $z=4.500$ and $4.667$, respectively.
Since the $i$ and $z$ observations do not suffer from strong absorption and there is no evidence for the presence of a flare at that time, we adopt them to construct the intrinsic optical ($z$ band) to X-ray SED to be $f_\nu \propto \nu^{-0.70\pm0.05}$, with which we can obtain the absorption correction in $r,~g$ as well as UVOT White and V bands. In the direction of GRB 220101A,  the Galactic extinction is $E(B-V)=0.0483$\cite{Schlafly2011}. Basing on the intrinsic spectrum of and assuming no extinction from the GRB host galaxy, we find an absorption in White band as high as $A_{\rm Wh}=4.78\pm0.10$ mag, including Lyman absorption and the Galactic extinction, see the bottom panel of Extended data Fig. 4.
Note that here  the central frequency of the White band observation has been taken as the same as that of the V band because of the serious absorption of the bluer photons, as demonstrated in Supplementary Fig. 1. 

In this work we adopt a cosmology with with H$_{0}=$ 67.4 km s$^{-1}$  Mpc $^{-1}$, $\Omega_{\rm M} =0.315$ and 
$\Omega_{\Lambda}  = 0.685$\cite{Planck2020}, a redshift $z=$4.618 leads to a distance modules DM$=48.19$. The absolute peak magnitude is calculated via $M_{\rm peak,abs}=M_{\rm peak}-{\rm DM}-A_{\lambda}+2.5(1-\beta_{\rm i})\log(1+z)$, where the last term is the k-correction and $\beta_{\rm i}$ is the intrinsic spectral slope. The pity is that none of the extremely luminous flashes in GRB 990123, GRB 050904, GRB 080319B and GRB 220101A have a measured optical/ultraviolet spectrum. For GRB 220101A, the UVOT and XRT data suggest an ``overall" optical to soft X-ray spectrum softer than $\nu^{-1.3}$. If this holds in the optical band (i.e., $\beta_{\rm i}\geq 1.3$), then we would have $M_{\rm peak,abs}\leq -40$ mag in the visible band. Though suffering from some uncertainty, it is so far the unique candidate to be brighter than the absolute magnitude of $-39$ mag. In Fig. 3 we have assumed $\beta_{\rm i}=1$, as usual, for all bursts.

{\bf The numerical interpretation of the optical emission as well as the X-ray afterglow emission}. Here we define the X-ray emission after $\sim 170$ s after the {\it Swift} trigger as the afterglow since the earlier emission are most likely the low energy part of the prompt radiation arising from the internal energy dissipation. 

{\it The extremely-bright optical flare induced by the interaction of late-ejected fast/main ejecta with the early outflow.}
In the prompt gamma-ray emission lightcurve, there are several weak gamma-ray spikes from earlier outflow before the main pulse starting at $\sim T_0 + 65$ s.
The front half part (between $\sim T_0 + 65$ to 102 s) of the giant gamma-ray pulse should come from the energy release of the main outflow, 
either dissipated through internal shocks or magnetic re-connections within it.
For the later part ($> T_0 + 102$ s) of the giant pulse, it overlaps with an energetic optical/ultraviolet flare,
which indicates the rise of an additional dissipation process. 
As the preceding weak outflow gets decelerated to a bulk Lorentz factor of $\Gamma_1$, 
a later launched but faster shell (with a bulk Lorentz factor of $\Gamma_4$) will catch up with it at a radius of $R_0$,
so that a collision between two shells would occur. 
Note that $\Gamma_1$ and $\Gamma_4$ correspond to $\Gamma_{\rm W}$ and $\Gamma_{\rm M}$ mentioned in the main text,
which is used here for the convenience of the discussion below.
If the fast shell is not extremely magnetized, the collision would produce a forward shock (FS) propagating into materials of the preceding shell,
and a reverse shock (RS) propagating into the fast shell.
As a result, an optical/ultraviolet flash is expected from the radiation in the downstream of the RS,
which has been initially proposed and works well for optical flash in the early afterglow stage\cite{Sari1999}.
Here we show that this scenario could account for the prompt optical emission of GRB 220101A with a detailed numerical approach.

Two shocks separate the system into four regions: 
(1) the unshocked slow shell, (2) the shocked slow shell, (3) the shocked fast shell, and (4) the unshocked fast shell.
Hereafter, $X_i$ denotes the value of the quantity $X$ in Region ``$i$'' in its own rest frame.
Unlike the preceding shell that exhausts the magnetic energy in the early stage ($\sigma_1$ = 0),
the later fast shell may keep the magnetic fields advected from the central engine,
which could be parameterized by the magnetization of $\sigma_{4} = {B^{2}_{4}}/{4 \pi n_4 m_{\rm p} c^2}$,
where $n_4$ is the particle density in the comoving frame of Region 4 and $m_{\rm p}$ is the proton mass.
Let's introduce an equivalent ``luminosity'' of the kinetic, internal and the magnetic energy
for the two shells measured in the lab frame, $L_i$, 
the corresponding particle density is then
$n_{i} = {L_i}/{4 \pi R^2 \beta_i \Gamma_i^2 m_{\rm p} c^3 (1+\sigma_i)}$, 
where $i = 1,4$, $\beta_i = 1/\sqrt{(1-1/\Gamma_i^2)}$ and $R$ is the radius from the central engine.
Due to the highly variable nature of the outflow from the central engine,
the luminosity of the later fast ejecta could be further described by
$L_4 =L_{\rm f} (R/R_{\rm peak})^{q_{\rm r}}$ for $R \le R_{\rm peak}$
and $L_4=
L_{\rm f} (R/R_{\rm peak})^{q_{\rm d}}$  for  $R > R_{\rm peak}$, 
where $R_{\rm peak}$ is the radius that the RS reaches its peak luminosity,
and $q_{\rm r}$ ($q_{\rm d}$) is the rising (decaying) index of the luminosity before(after) $R_{\rm peak}$.
We assume that Region 2 and Region 3 share a common bulk Lorentz factor ($\Gamma_2 = \Gamma_3$).
After applying the hydrodynamical/magnetohydrodynamical jump conditions \cite{Fan04a,Zhang05} to the FS/RS respectively
and the energy conservation law to the FS-RS system\cite{Geng16},
the evolution of $\Gamma_2$ and relevant quantities within these regions could be solved numerically
given the total isotropic energy of each shell ($E_{\rm f}$ and $E_{\rm s}$).

The kinetic particle-in-cell simulations reveal that particle acceleration is less efficient
in strongly magnetized shock than that of weakly magnetized shock\cite{Sironi15}.
The shock is considered to be moderately magnetized, and it is reasonable to assume that the distribution of electrons injected downstream 
is Maxwellian both for the FS/RS\cite{Giannios09}, i.e., 
$Q_i (\gamma_{\rm e},t) = 
C_i \left({\gamma_{\rm e}}/{\gamma_{\mathrm{c},i}} \right)^2 \exp^{-\gamma_{\rm e}/\gamma_{\mathrm{c},i}}$,
where $\gamma_{\mathrm{c},i} = \frac{1}{3} \epsilon_{\mathrm{e},i} \frac{e_i}{\rho_i c^2} \frac{m_p}{m_e}$ is the typical Lorentz factor of the thermal distribution,
$\epsilon_{\mathrm{e},i}$ is the fraction of post-shock energy that goes into electrons for each region,
$e_i$ and $\rho_i$ is the energy and density of protons.
The normalization constant $C_i$ is obtained from the relevant mass conservation.
The instantaneous electron spectrum can be obtained by solving the continuity equation of electrons in energy space\cite{Geng18}.
Due to the overlap of the prompt gamma-rays with the optical/ultraviolet radiation region,
the inverse Compton cooling of electrons is significant and hence has been properly incorporated in our calculation. This makes the calculation very time-consuming and it is not conventional to explore in a large parameter space. 
Integrating the synchrotron radiation power from the electron spectrum in Regions 2 and 3
and considering the effect of synchrotron self-absorption and the equal-arrival-time surface, 
the radiation spectra and the light curves are then derived. 
With a starting radius of $R_0 = 10^{15}$~cm for the collision
and a set of parameters of $L_1 \sim 5.6 \times 10^{52}$~erg~s$^{-1}$, $L_{\rm f} \sim 4.5 \times 10^{53}$~erg~s$^{-1}$, 
$\Gamma_1 \sim 100$, $\Gamma_4 \sim 1000$, $q_r \sim 1.3$, $q_d \sim -0.5$,
$\sigma_4 \sim 0.1$, $\epsilon_{B,3} \sim 0.08$, $\epsilon_{\mathrm{e},2} \sim 0.1$, $\epsilon_{\mathrm{e},3} \sim 0.07$,
$E_{\rm s} \sim 5.8 \times 10^{53}$~erg, $E_{\rm f} \sim 6.0 \times 10^{54}$~erg, 
our numerical optical lightcurves are consistent with the observation data prior to 170s after the BAT trigger, as shown in Extended Data Fig.5. 
Please bear in mind that the above calculation is just to show that the data can be reproduced within the current framework and the physical parameters are not strictly fitted.

{\it The external forward and reverse shock emission.} In our modeling, it turns out that the shallow-declining part of the optical flare is hard to be accounted for (see Extended data Fig. 5). A possibility is the emergence of the reverse shock, as observed in for instance GRB 990123\cite{Akerlof1999,Sari1999}. Indeed, a reverse and forward shock scenario can reasonably reproduce the optical and X-ray data. The magnetic field in the reverse shock region should be stronger than that in the forward shock region by a factor of quite a few $\times 10$ otherwise the induced optical flash can not be  brighter than the forward shock peak optical emission by a factor of $\sim 1000$ \cite{Fan2002,Zhang2003}. The following physical parameters are found to be able to reasonably reproduce the afterglow data:  the isotropic energy is $E_{\rm iso}=1.0\times10^{\rm 55}$ erg with a half open jet angle $\theta_{\rm j}=0.025$, the initial Lorentz factor is $\Gamma=800$,  the fraction of forward and reverse shock energy given to the electrons is $\epsilon_{\rm e}=0.4$, the fraction of the forward (reverse) shock energy given to the magnetic field is $\epsilon_{\rm b,fs}=2.5\times 10^{-5}$ ($\epsilon_{\rm b,rs}=0.3$), the number density of the interstellar medium is $n=0.05~{\rm cm^{-3}}$ and the power-law index for shock-accelerated electrons is $p=2.26$. Such a $p$ is well consistent with that needed in reproducing the optical to X-ray spectrum and lightcurves shown in Extended data Fig. 4 and Extended data Fig. 5,  including {\it Swift} data analyzed in this work and Liverpool telescope data from GCN\cite{2022GCN.31357....1P,2022GCN.31425....1P}.

{\bf The prospect of detecting ultra-luminous optical/ultraviolet flashes at high redshifts with SVOM/VT}.  Optical/ultraviolet flashes at high redshift will suffer from serious absorption.  Following Moller \& Jakobsen\cite{Moller1990}, we estimate the absorption correction to be A$_{\rm B} \sim 5$ mag (the received photons are mainly caused by red leak of blue filter) and A$_{\rm R} \sim 1$ mag for the sources at $z\sim 6$, based on the responses of SVOM/VT blue and red channels (i.e., B and R). For flashes as luminous as GRB 080319B or GRB 220101A, if taken place at $z\sim 6$, then we would have $M_{\rm R}\sim 10.5$ mag and $M_{\rm B}\sim 15$ mag. With the shortest exposure of $1$s, SVOM/VT has a  dynamic range of $9-18$ mag, which is sufficiently sensitive to catch the signals mentioned above. 
However, usually the exposure time of SVOM/VT should be 10-100 seconds, for which the R filter may get saturated but the B filter is not.   
We therefore conclude that SVOM/VT is a suitable equipment to detect the extremely bright optical flares of GRBs at $z\sim6$.

\end{methods}

\begin{addendum}


\item[Data availability.] The  {\it Swift} and Fermi data analysed/used in this work are all publicly available \\
at https://heasarc.gsfc.nasa.gov/cgi-bin/W3Browse/.

\item[Code availability.] HEASoft code is available at https://heasarc.gsfc.nasa.gov/lheasoft/ and calibration database ({\it CALDB}) is available at https://heasarc.gsfc.nasa.gov/docs/heasarc/caldb/swift/. Fermi {\it GBM Data Tools} is available at 
https://fermi.gsfc.nasa.gov/ssc/data/analysis/gbm. {\it Bilby} is available at https://git.
ligo.org/lscsoft/bilby/.

\item[Acknowledgements] This work was supported in part by NSFC under grants of No. 12225305, 11921003, 11933010 and 12273113, the China Manned Space Project (NO.CMS-CSST-2021-A13), Major Science and Technology Project of Qinghai Province  (2019-ZJ-A10), Key Research Program of Frontier Sciences (No. QYZDJ-SSW-SYS024). JJG has been supported by the Youth Innovation Promotion Association (2023331). SC has been supported by ASI grant I/004/11/0.

\item[Author Contributions] Y.Z.F and Z.P.J launched the project. Z.P.J, H.Z., Y.W, X.L, S.C and J.Y.W carried out the data analysis. Y.Z.F, J.J.G., X.F.W, D.M.W and Z.P.J interpreted the data.  Z.P.J, H.Z. and Y.Z.F prepared the paper and all authors joined the discussion. Z.P.J and H.Z contributed equally. 

\item[Competing interests] The authors declare no competing interests.

\item[Additional information]

{\bf Correspondence and requests for materials} should be addressed to Y.Z.F (yzfan@pmo.ac.cn).

\end{addendum}


\begin{figure}[ht]
\label{fig:figure1}
\begin{center}
\includegraphics[width=0.99\columnwidth]{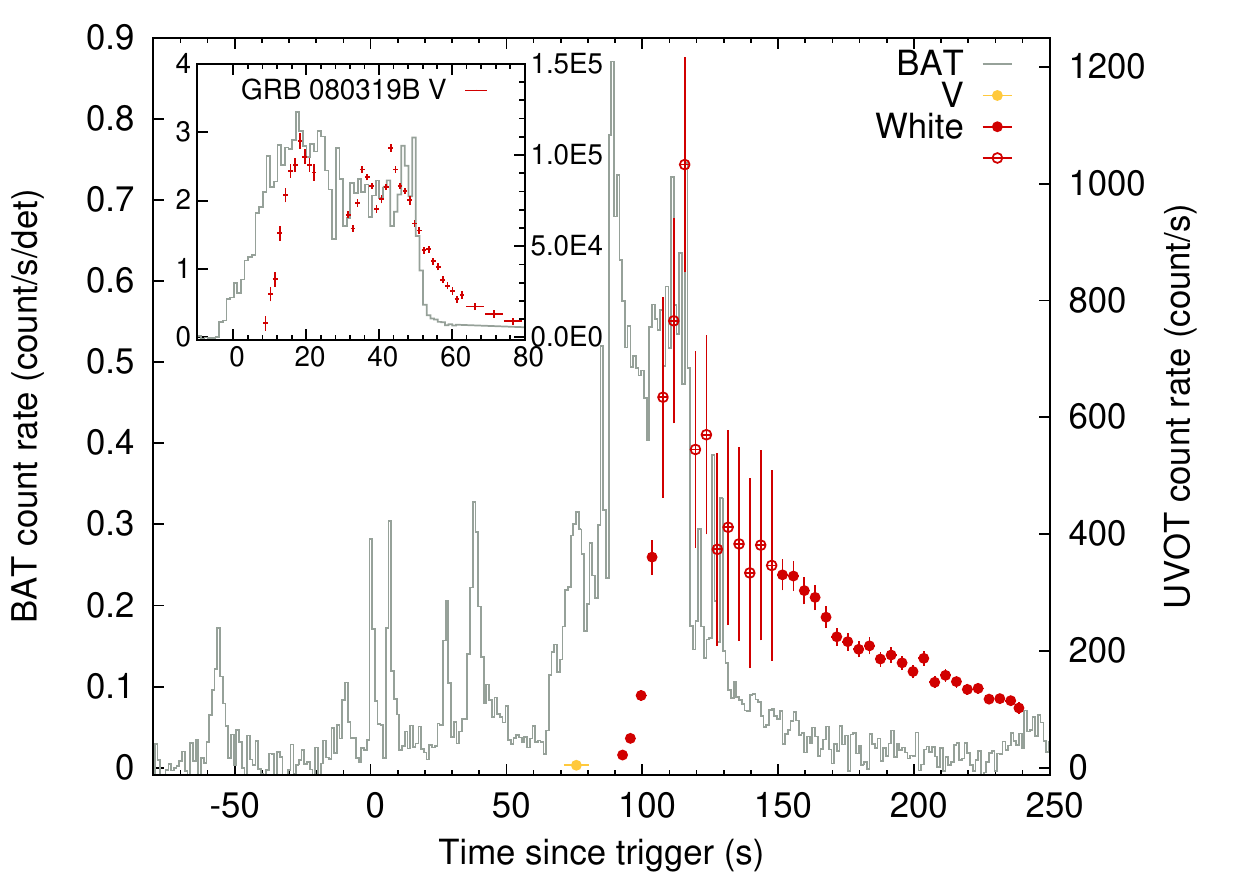}
\end{center}
\caption{
{\bf Photon count rates of the prompt gamma-ray ({\it Swift}/BAT) and optical ({\it Swift}/UVOT V and White band) emission of GRB 220101A.} The prompt gamma-ray lightcurve is highly variable, while the prompt optical emission lightcurve is relatively smooth and does not trace that of gamma-rays.  The red filled circles are from the aperture measurement while the open circles are obtained with the new method developed in this work. The energetic optical/ultraviolet flare just overlaps with the late part of the giant outburst phase of the prompt gamma-rays. In the early time,  the gamma-ray emission consists of a few weak spikes and the corresponding outflow got decelerated by the interstellar medium. The much more energetic outflow giving rise to the gamma-ray emission in the giant phase is expected to move with a very high Lorentz factor and the collision/catching-up with the earlier/decelerated outflow at a radius of $\sim 10^{16}$ cm would generate strong shocks and hence extremely bright ultraviolet/optical emission.The error bars represent the $1\sigma$ statistical errors.
}
\end{figure}

\begin{figure}[ht]
\begin{center}
\includegraphics[width=0.99 \columnwidth]{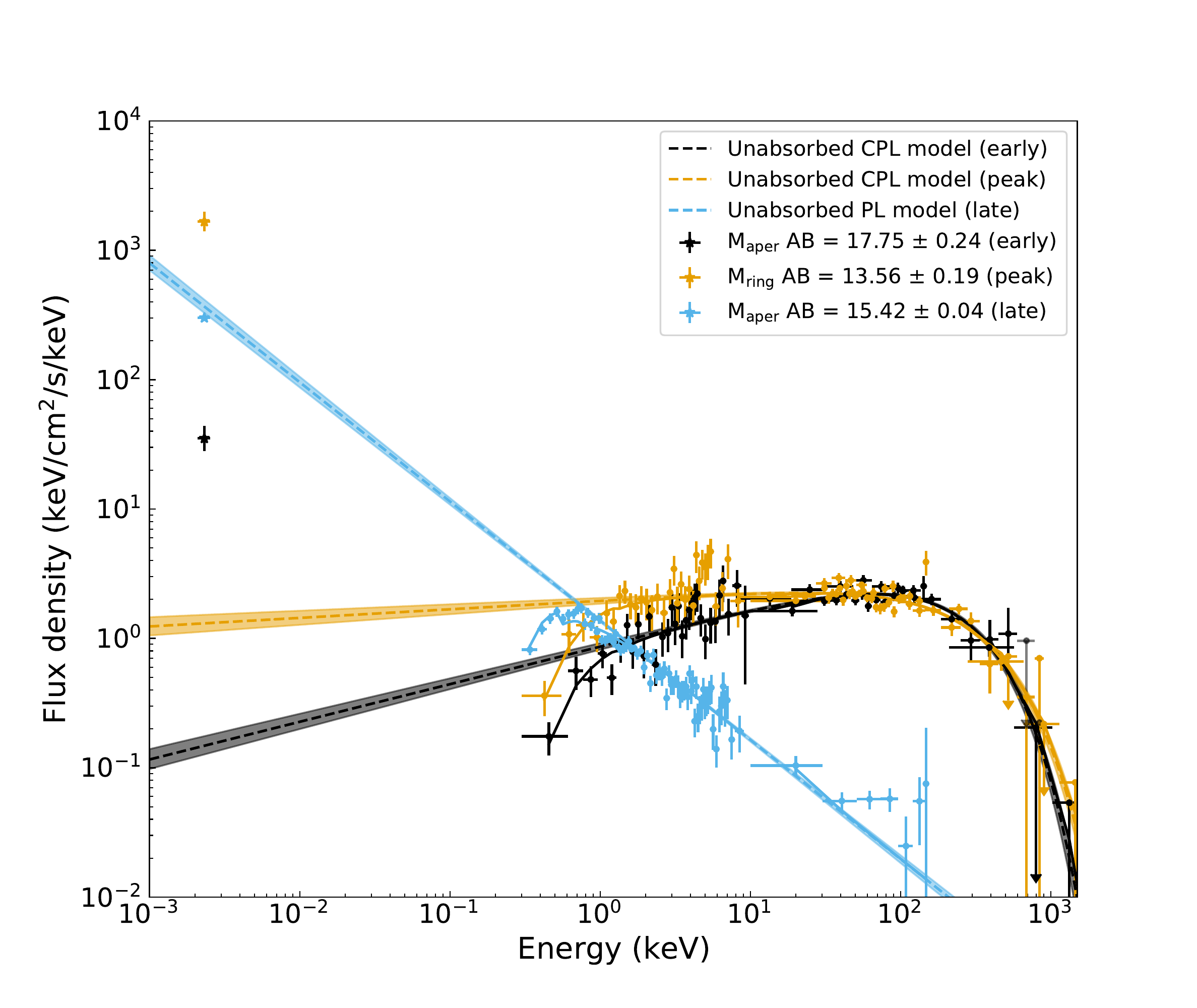}
\end{center}
\caption{
{\bf  The ``prompt" optical to gamma-ray SEDs of GRB 220101A.} 
The data in black are collected at the early stage ($91.96 - 93.62$ seconds) of the UVOT/White band measurement. The data in orange are for the peak time ($113.64-117.62$ seconds) of the UVOT/White emission. Cutoff power-law (CPL) fits to X-ray and gamma-ray at these two epochs are shown in dashed lines. The joint statistics over degrees of freedom (Stat/d.o.f) for the fits are 505.8/610 and 630.6/615, respectively. For both measurements, the optical emission are well above the extrapolation of the high energy spectrum, suggesting an origin different from the prompt X-rays and gamma-rays. While in the late time interval of $t\sim 173.6-239.6$ seconds, the extrapolation of power-law (PL) fit to the X-ray and gamma-ray spectrum (with Stat/d.o.f=638.7/638) into the optical is in agreement with the UVOT data. Since the temporal decline slope of the UVOT data is also shallower than that in the time interval of $117.62-173.63$ seconds, we suggest that they likely consist of also the reverse shock emission with a soft spectrum and the softening of the X-ray spectrum is because of the emergence of this new component. The error bars represent the $1\sigma$ statistical errors and the upper limits are at the 3$\sigma$ confidence level.
} 
\label{fig:figure2}
\end{figure}

\begin{figure}[ht]
\begin{center}
\includegraphics[width=0.99 \columnwidth]{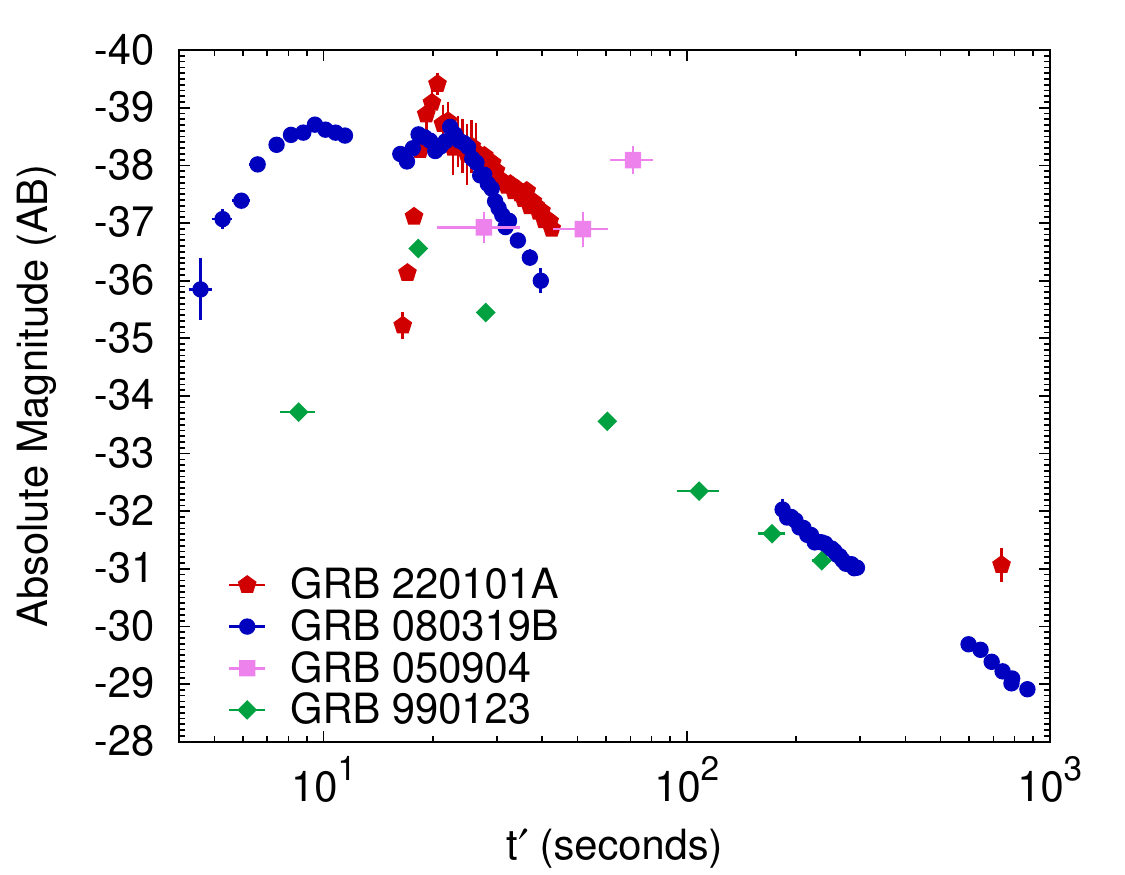}
\end{center}
\caption{
{\bf  The ultraviolet/optical flare of GRB 220101A in comparison to some other extremely energetic optical flares of GRBs.}
The White band emission of GRB 220101A has been corrected for total extinction of $A_{\rm Wh}=4.78\pm 0.10$ mag, including the tiny softening of $E(B-V)=0.0483$ mag in the Milky Way. The absolute AB magnitude of GRB 220101A (red) exceeds that of GRB 080319B (blue) \cite{Racusin2008}, the so-called naked burst, as well as GRB 990123 (green) \cite{Akerlof1999} and GRB 050904 (pink)\cite{Boer2006}, rendering it the most energetic optical/ultraviolet flare recorded so far. More information for comparison, such as the redshifts, bands and extinctions, are listed in Supplementary Table 6. The error bars represent the $1\sigma$ statistical errors.
}
\label{fig:figure3}
\end{figure}

\renewcommand{\figurename}{Extended data Figure} 
\setcounter{figure}{0} 

\clearpage
\begin{figure}[htbp]
\begin{center}
\includegraphics[width=0.9\textwidth]{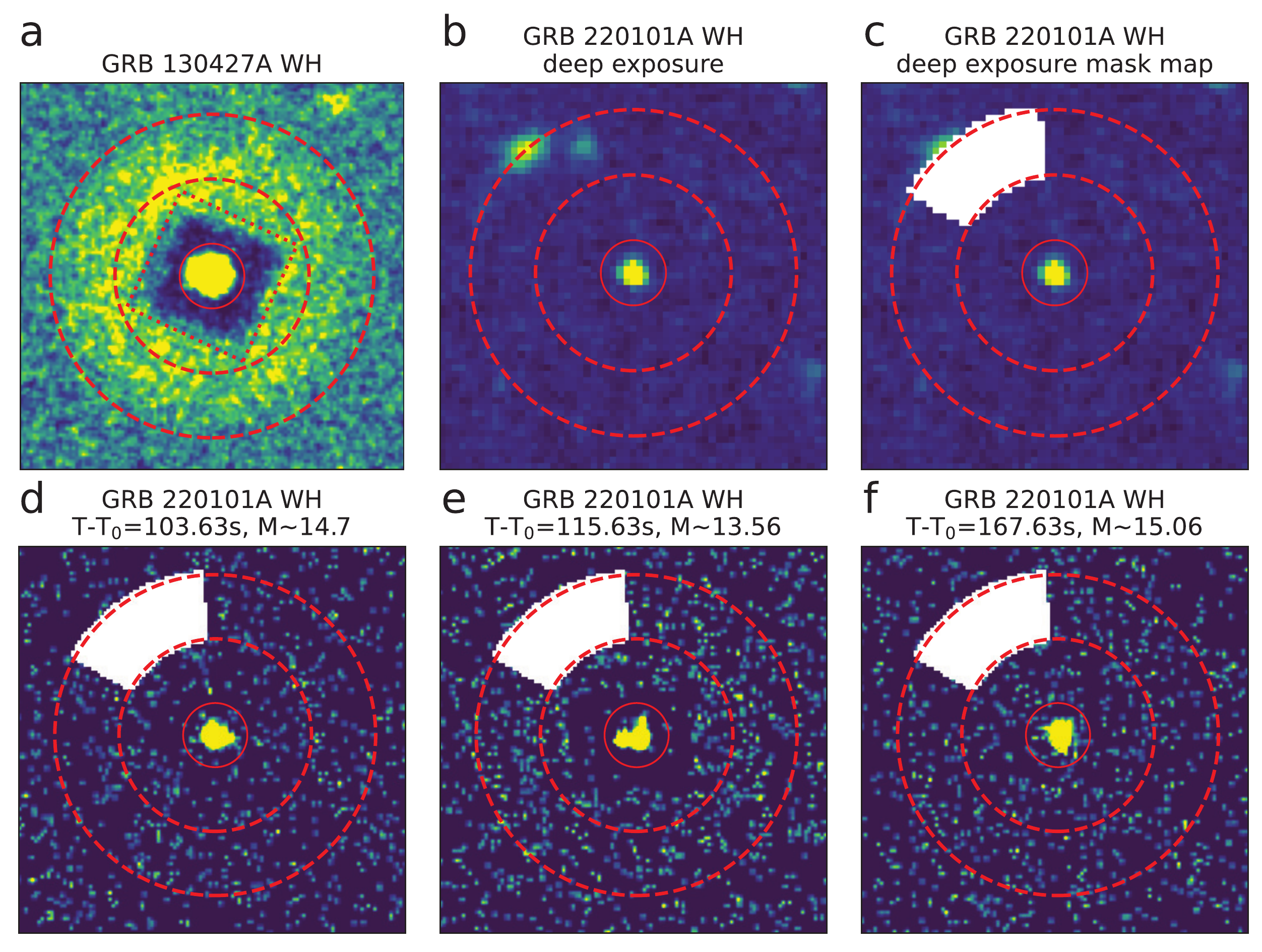}
\end{center}
\caption{
{\bf {\it Swift}/UVOT white (WH) band images demonstrating the halo ring photometry method.} 
Panel (a) is the White band image of GRB 130427A, where the solid circle represents the standard aperture of UVOT with a radius of 5 arcsec. The dotted square region strongly suffered from coincidence loss with a typical side length of $\sim$ 20 arcsec. Dashed annulus with an inner radius of 15 arcsec and an out radius of 25 arcsec is the halo ring region defined in this work, for which the $\dot{N}_{\rm ring}$ is derived. Panel (b) shows the deep exposure of GRB 220101A field in White band, which reveals 2 faint sources in the halo ring region, hence we masked the annulus region from 95$^\circ$ to 150$^\circ$, as shown in panel (c). In addition, images of panel (b) and (c) have a pixel scale of 1.004 arcsec/pixel instead of 0.502 arcsec/pixel for other 4 images. Panel (d), (e) and (f) show some images 
around the peak time of GRB 220101A. We measured count rates in unmasked annulus region and corrected it to the whole annulus region.
}
\label{fig:images}
\end{figure}

\clearpage
\begin{figure}[htbp]
\begin{center}
\includegraphics[width=0.9\textwidth]{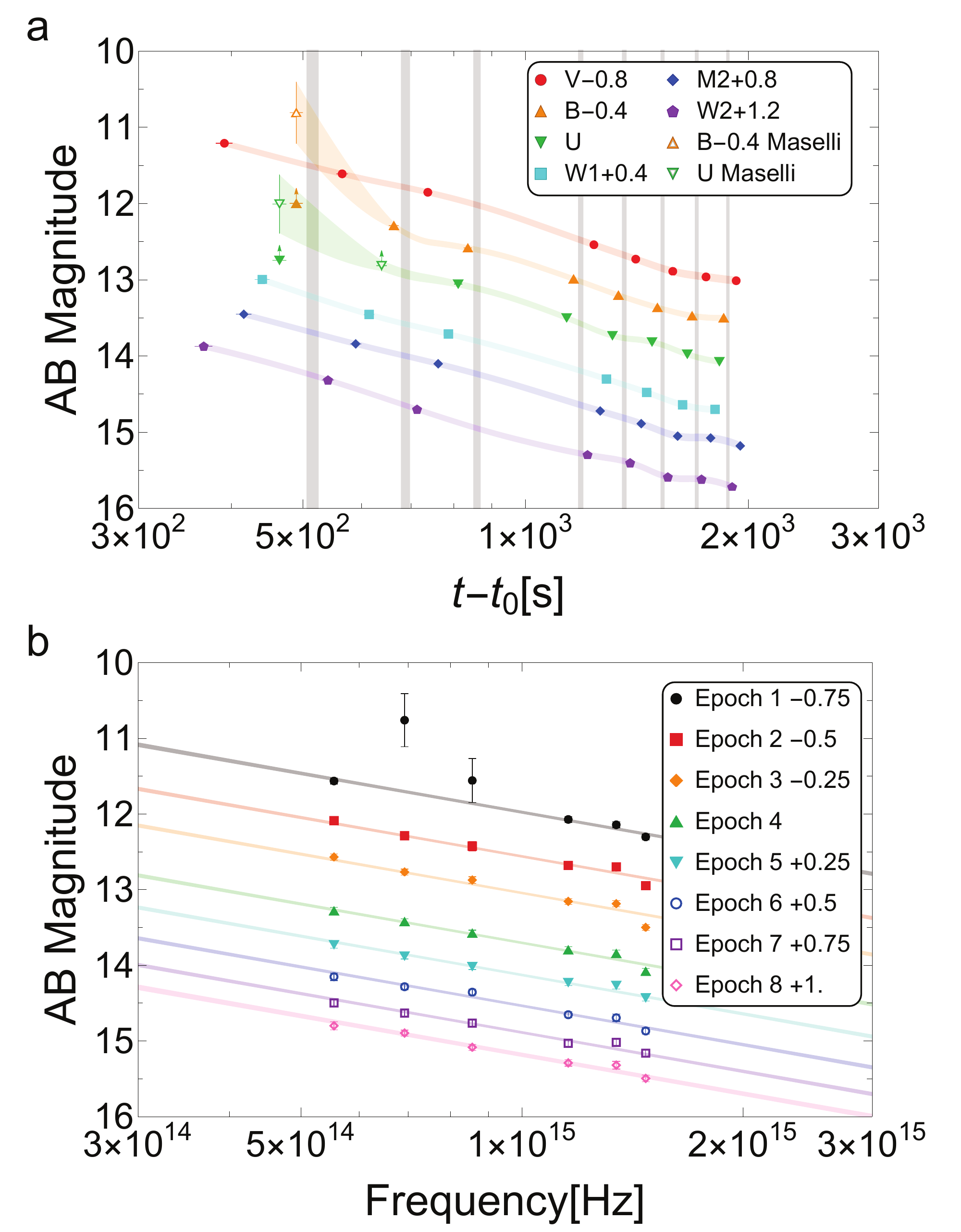}
\end{center}
\caption{
{\bf The UVOT lightcurves (left) as well as the SEDs (right) of GRB 130427A.} In the top panel, the vertical grey regions mark the observation periods of the White filter. Note that the second U-band data is saturated, which was however a detection point in Maselli et al.\cite{Maselli2014} 
if only event data in the last 6s was measured, hence the filled and the empty green triangles coincide. The shaded colorful regions across photometry points are our interpolation results of light curve.
The bottom panel presents the optical to ultraviolet SEDs at the White band observation times constructed with the extrapolated UVOT narrow band data.  
A single power-law spectrum well reproduces the data, as anticipated in the fireball external forward shock afterglow model, with which a reliable evaluation of the White band emission is yielded, as reported in the last column of Supplementary Table \ref{tab:130427Awh}.
}
\label{fig:GRB130427_SED}
\end{figure}

\clearpage
\begin{figure}[htbp]
\begin{center}
\includegraphics[width=0.9\textwidth]{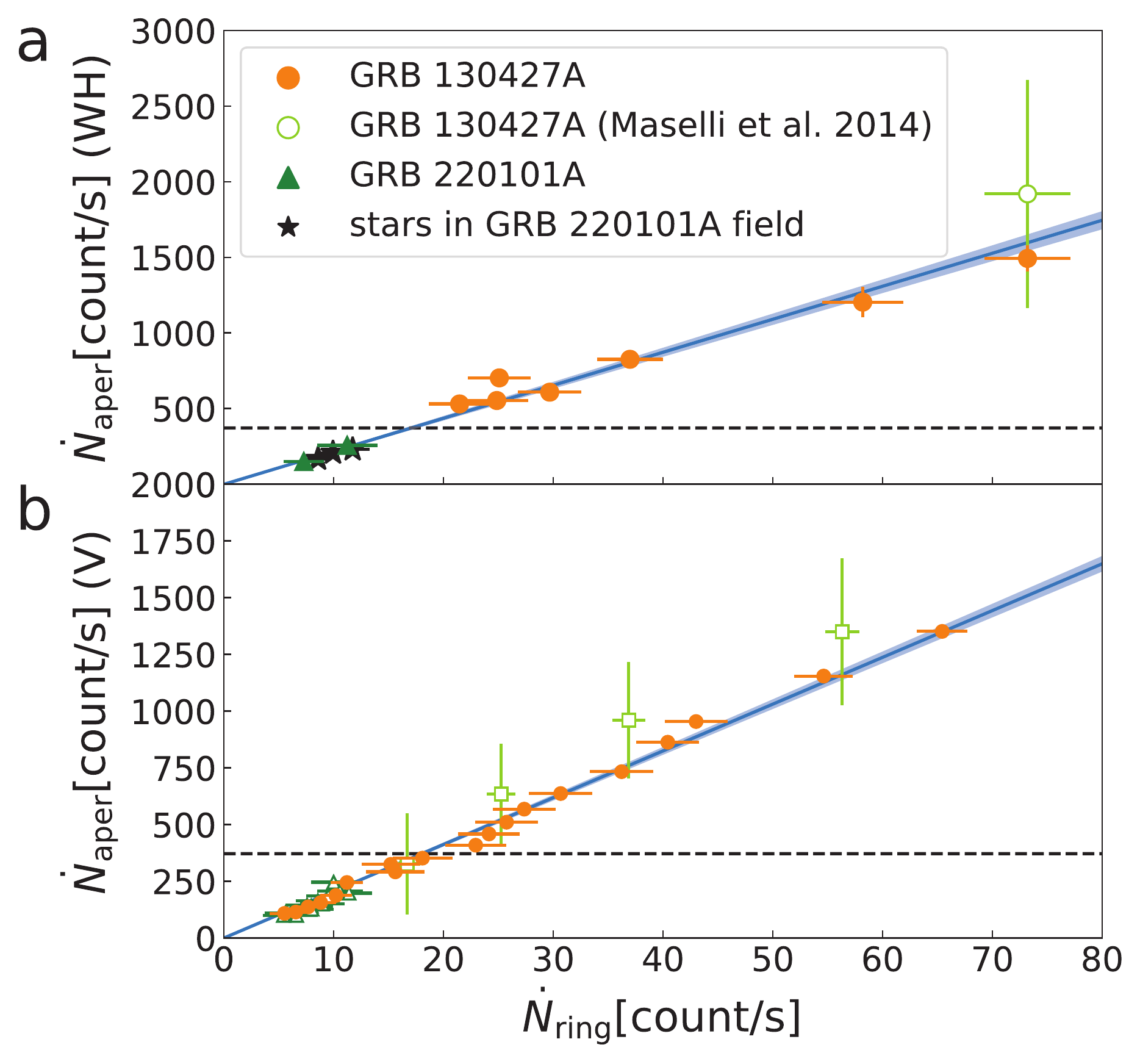}
\end{center}
\caption{
{\bf  Photon count rates in $5''$ aperture $\dot{N}_{\rm aper}$ (directly measured or inferred from the intrinsic value $\dot{N}_{\rm int}$) and $15-25''$ ring $\dot{N}_{\rm ring}$ (coincidence loss corrected) for some {\it Swift}/UVOT white and V band measurements.} The top panel is for the White band. The dark green upward triangles represent the two unsaturated measurements of GRB 220101A in the tail phase of the flare. The filled squares are for three bright stars in the filed of GRB 220101A. The light green downward empty triangle represents inferred $\dot{N}_{\rm aper}$ with the photometry result of GRB 130427A derived with readout streak method\cite{Maselli2014}. 
As for orange points, the vertical coordinate represents the White-band emission of GRB 130427A inferred from measurements in other UVOT bands (see Extended data Fig. \ref{fig:GRB130427_SED}), while the horizontal coordinate is the $\dot{N}_{\rm ring}$ (see Extended data Fig. \ref{fig:images} for definition). Black squares are 3 unsaturated stars in GRB 220101A field. The bottom panel is for the V band. Empty dark green triangles are unsaturated measurements of GRB 080319B with \textit{HEASoft} and empty light green squares are photometry results of GRB 080319B derived with the readout streak method\cite{Page2013}. As for orange points, the vertical coordinate represents the photometry result of GRB 080319B observed with RAPTOR-T\cite{2009ApJ...691..495W} when the UVOT observations were ongoing, while the horizontal axis represents $\dot{N}_{\rm ring}$ in the corresponding UVOT V-band image. The linear fit is just for filled points in both panels, and the correlation coefficients of filled points are 0.990 and 0.998 for the left and bottom panel, respectively. Black dashed lines represent the saturation count rate (coincidence loss corrected, $\sim372$ count s$^{-1}$) of UVOT.
}
\label{fig:apering}
\end{figure}

\clearpage
\begin{figure}[htbp]
\begin{center}
\includegraphics[width=0.9 \columnwidth]{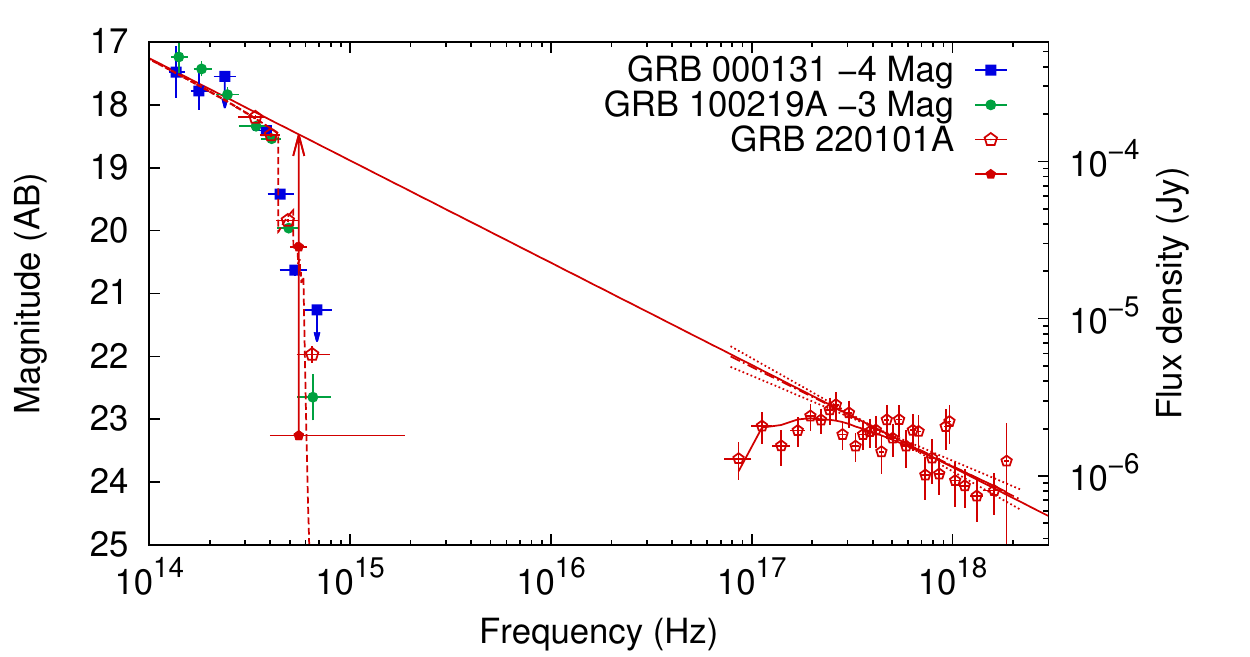}
\end{center}
\caption{
{\bf Optical to X-ray SED of GRB 220101A.}
{{\it Swift} XRT, UVOT and $g,~r,~i,~z$ observations of Liverpool telescope in the time interval of $t\sim 0.62-0.68$ day after the burst\cite{2022GCN.31357....1P}. Such a set of ground-based telescope observation data are chosen because they are almost simultaneous with one UVOT White exposure. Neither the X-ray nor the optical emission displays a flare. Therefore, we construct the optical SED with the data collected at $t\sim 0.625$ day. We find that the absorption correction is $A_{\rm Wh}=4.78$ mag for intrinsic optical to X-ray spectrum with index $\beta_{\rm oX}=0.65$, it is well consistent with X-ray spectrum $\beta_{\rm X}=0.63\pm0.09$. The central frequency of the White band observation has been taken as the same as that of the V band because of the serious absorption of the bluer photons, as demonstrated in Extended data Fig. \ref{fig:equivalent}. The optical SEDs of other two GRBs 000131\cite{Andersen2000} and 100219A\cite{Thone2013} at similar redshifts ($z=4.500$ and $4.667$, respectively) are also shown for comparison.
}
}
\label{fig:SED}
\end{figure}

\begin{figure}[htbp]
\begin{center}
\includegraphics[width=0.9\columnwidth]{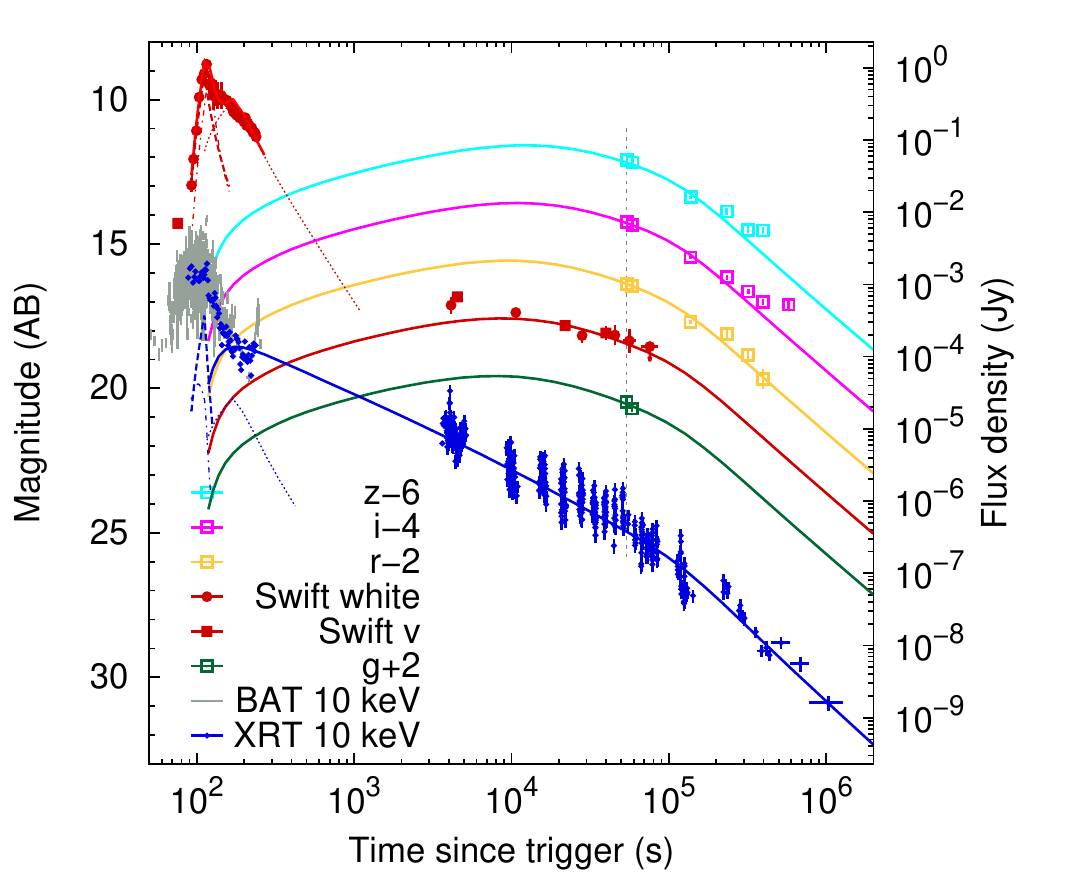}
\end{center}
\caption{
{\bf Fit to the multi-band afterglow lightcurves of GRB 220101A.}
\label{fig:AfterglowFit}
The {\it Swift} XRT and UVOT data are analyzed in this work, 
and the other optical data are adopted from Liverpool telescope \cite{2022GCN.31357....1P,2022GCN.31425....1P}.  
The total extinction corrections, including Galactic extinction and interstellar-medium extinction are $A_{\rm Wh}=4.78$, $A_{\rm v}=1.88$, $A_{g}=3.51$, $A_{r}=1.46$, $A_{i}=0.24$ and $A_{z}=0.10$, respectively. 
The dashed and dash-dotted lines represents forward and reverse shock emission arising from the weak/slow and main/fast outflow collision. Solid and dotted lines are the regular external forward and reverse shock emission of the outflow. In our calculation, the main/fast outflow was launched 92 seconds after the BAT trigger. 
Note that the X-ray emission at $t\leq 170$ sec was attributed to the low energy part of the prompt emission and has not been addressed in our modeling.} 
\label{fig:lightcurve-fitting}
\end{figure}

\clearpage
\begin{figure}[htbp]
\begin{center}
\includegraphics[width=0.9\columnwidth]{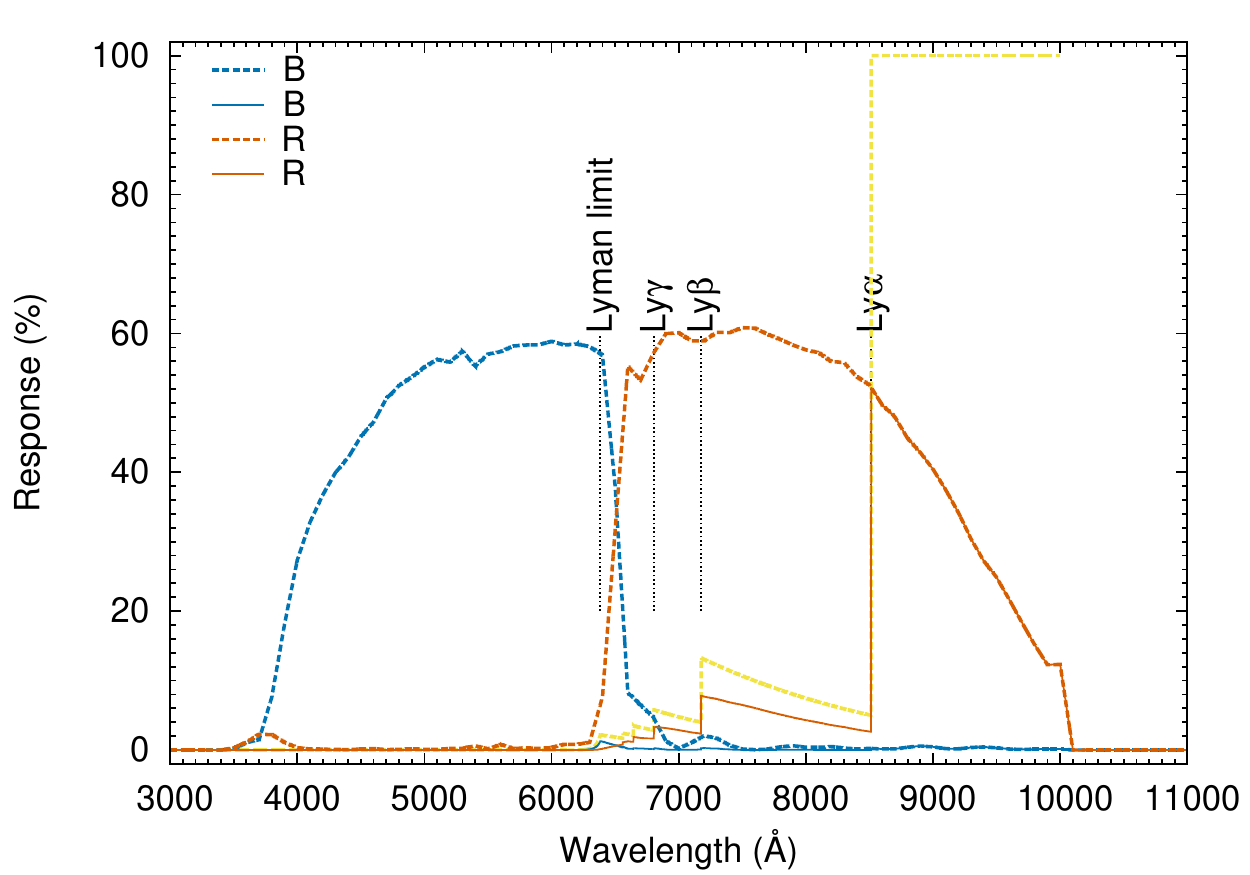}
\end{center}
\caption{{\bf The response of the SVOM/VT and the Lyman absorption of the high redshift ($\sim 6$) event.} 
The optical/ultraviolet flash will surfer from strong absorption by intergalactic medium. Following Moller \& Jakobsen\cite{Moller1990}, we find that A$_{\rm B} \sim 5$ mag (the received photons are mainly caused by red leak of blue filter) and A$_{\rm R} \sim 1$ mag for a source at $z=6$, based on the responses of SVOM/VT blue and red channels (i.e., B and R). If the initial flash is as bright as that detected in GRB 080319B and GRB 220101A, the absorbed one would still be caught by SVOM/VT with a dynamic range of $9-18$ mag for the shortest exposure of 1s. Therefore SVOM/VT is a suitable equipment to detected extremely bright optical flares of GRBs at $z\sim6$.
}
\label{fig:SVOM}
\end{figure}


\renewcommand{\tablename}{Supplementary Table} 
\captionsetup[table]{position=above}

\clearpage
\section*{Supplementary Information}

\renewcommand{\figurename}{Supplementary Figure} 
\setcounter{figure}{0} 

\begin{figure}[hbp]
\begin{center}
\includegraphics[width=0.9\columnwidth]{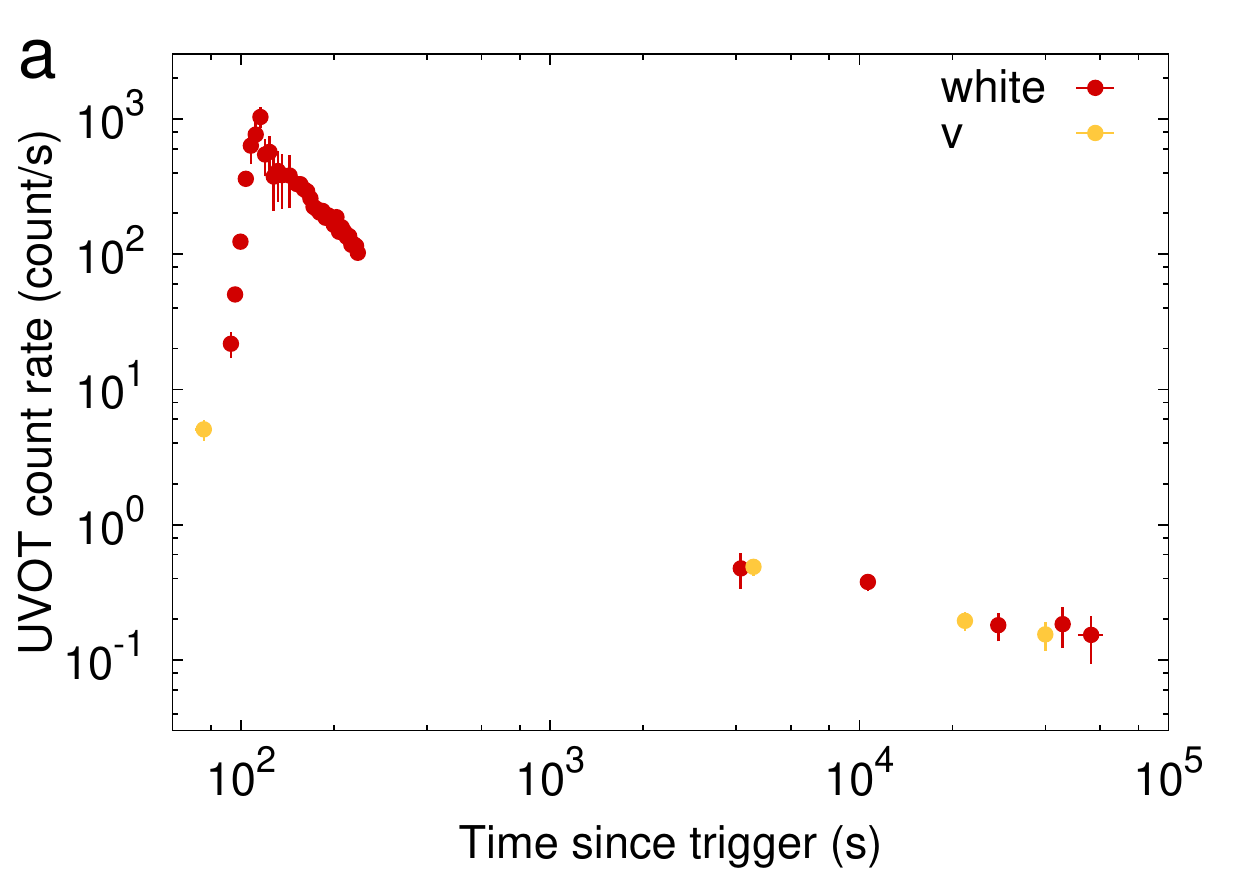}
\includegraphics[width=0.9\columnwidth]{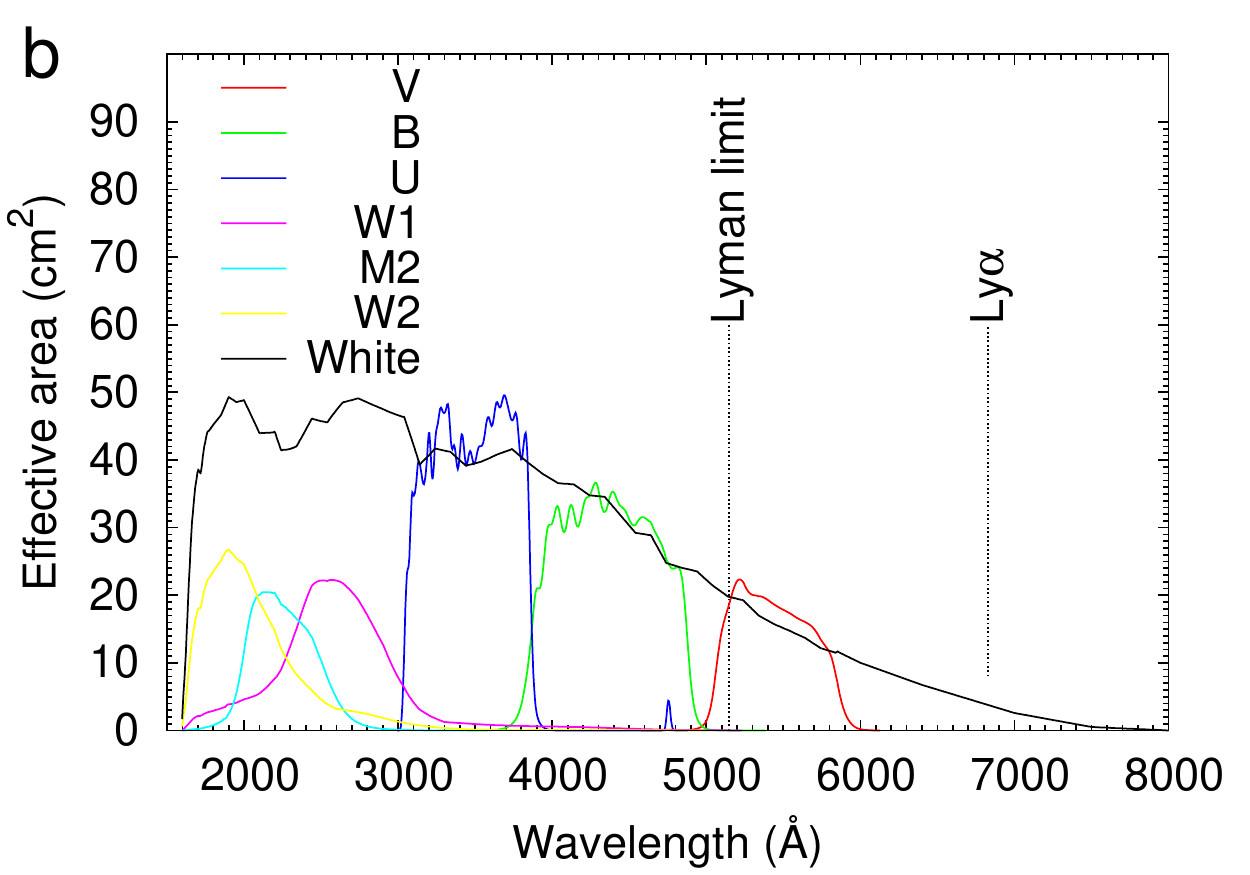}
\end{center}
\caption{{\bf The similarity of {\it Swift}/UVOT White and V band observations of GRB 220101A.}  
The top panel shows that the photon count rates in White band are almost the same as that in V band. This is because the photons with wavelengths below the Lyman limit (in the observer frame, it is  $5124$\AA; see the bottom panel) are almost fully absorbed, and the photons near the Lyman $\alpha$ may also suffer from strong absorption (see Extended data Fig. \ref{fig:SED} for this effect). Therefore the collected photons are mainly within the V band.
}
\label{fig:equivalent}
\end{figure}

\clearpage
\begin{table}[h!]
\caption{ {\bf Early observations of GRB 130427A by {\it Swift}-UVOT.} Galactic extinction $A_{\rm V}=0.055$, $A_{\rm B}=0.071$, $A_{\rm U}=0.087$, $A_{\rm W1}=0.118$, $A_{\rm M2}=0.163$ and $A_{\rm W2}=0.156$ have been applied. These data points have been plotted in the top panel of  Extended data Fig. \ref{fig:GRB130427_SED}.}
\begin{center}
\label{tab:130427A}
	\begin{tabular}{cccccccc}
		\hline
		\hline
		T-T$_0$ & Exp & V & B & U & W1 & M2 & W2 \\
    	(s) & (s) & (AB) & (AB) & (AB) & (AB) & (AB) & (AB)\\
		\hline
		\hline
		367.38 & 19.46 & ... & ... & ... & ... & ... & 12.67$\pm$0.04 \\
		391.76 & 19.45 & 12.01$\pm$0.04 & ... & ... & ... & ... & ... \\
		416.18 & 19.45 & ... & ... & ... & ... & 12.65$\pm$0.04 & ... \\
		440.84 & 19.44 & ... & ... & ... & 12.60$\pm$0.04 & ... & ... \\
		465.10 & 19.44 & ... & ... & 12.09$\pm$0.38$^a$ & ... & ... & ... \\
		490.09 & 19.45 & ... & 11.28$\pm$0.40$^a$ & ... & ... & ... & ... \\
		540.86 & 19.44 & ... & ... & ... & ... & ... & 13.12$\pm$0.04 \\
		565.28 & 19.40 & 12.41$\pm$0.04 & ... & ... & ... & ... & ... \\
		589.61 & 19.46 & ... & ... & ... & ... & 13.04$\pm$0.04 & ... \\
		614.82 & 19.44 & ... & ... & ... & 13.06$\pm$0.04 & ... & ... \\
		639.13 & 19.46 & ... & ... & 12.90$\pm$0.07$^a$ & ... & ... & ... \\
		663.96 & 19.46 & ... & 12.69$\pm$0.04 & ... & ... & ... & ... \\
		713.68 & 19.45 & ... & ... & ... & ... & ... & 13.50$\pm$0.04 \\
		737.97 & 19.44 & 12.65$\pm$0.04 & ... & ... & ... & ... & ... \\
		762.19 & 19.44 & ... & ... & ... & ... & 13.30$\pm$0.04 & ... \\
		786.88 & 19.44 & ... & ... & ... & 13.31$\pm$0.04 & ... & ... \\
		811.16 & 19.44 & ... & ... & 13.06$\pm$0.04 & ... & ... & ... \\
		835.91 & 19.45 & ... & 12.98$\pm$0.04 & ... & ... & ... & ... \\
		1136.89 & 19.45 & ... & ... & 13.50$\pm$0.04 & ... & ... & ... \\
		1161.73 & 19.46 & ... & 13.39$\pm$0.04 & ... & ... & ... & ... \\
		1213.17 & 19.44 & ... & ... & ... & ... & ... & 14.10$\pm$0.04 \\
		1237.51 & 19.44 & 13.34$\pm$0.05 & ... & ... & ... & ... & ... \\
		1261.91 & 19.43 & ... & ... & ... & ... & 13.92$\pm$0.05 & ... \\
		1286.75 & 19.44 & ... & ... & ... & 13.90$\pm$0.04 & ... & ... \\
		1311.01 & 19.45 & ... & ... & 13.73$\pm$0.04 & ... & ... & ... \\
		1335.68 & 19.44 & ... & 13.60$\pm$0.04 & ... & ... & ... & ... \\
		1385.28 & 19.40 & ... & ... & ... & ... & ... & 14.20$\pm$0.04 \\
		1409.66 & 19.43 & 13.53$\pm$0.05 & ... & ... & ... & ... & ... \\
		1433.98 & 19.45 & ... & ... & ... & ... & 14.09$\pm$0.05 & ... \\
		1458.64 & 19.44 & ... & ... & ... & 14.08$\pm$0.04 & ... & ... \\
		1482.87 & 19.44 & ... & ... & 13.82$\pm$0.04 & ... & ... & ... \\
		1508.07 & 19.45 & ... & 13.77$\pm$0.04 & ... & ... & ... & ... \\
		1557.68 & 19.44 & ... & ... & ... & ... & ... & 14.39$\pm$0.04 \\
		1581.95 & 19.45 & 13.69$\pm$0.05 & ... & ... & ... & ... & ... \\
		1606.20 & 19.44 & ... & ... & ... & ... & 14.25$\pm$0.05 & ... \\
		1630.88 & 19.45 & ... & ... & ... & 14.24$\pm$0.04 & ... & ... \\
		1655.08 & 19.41 & ... & ... & 13.98$\pm$0.04 & ... & ... & ... \\
		1679.93 & 19.44 & ... & 13.87$\pm$0.04 & ... & ... & ... & ... \\
		1729.85 & 19.44 & ... & ... & ... & ... & ... & 14.42$\pm$0.04 \\
		1754.32 & 19.46 & 13.76$\pm$0.05 & ... & ... & ... & ... & ... \\
		1779.91 & 19.55 & ... & ... & ... & ... & 14.28$\pm$0.05 & ... \\
		1804.55 & 19.45 & ... & ... & ... & 14.30$\pm$0.04 & ... & ... \\
		1828.75 & 19.45 & ... & ... & 14.08$\pm$0.04 & ... & ... & ... \\
		1853.48 & 19.45 & ... & 13.90$\pm$0.04 & ... & ... & ... & ... \\
		1903.00 & 19.44 & ... & ... & ... & ... & ... & 14.52$\pm$0.05 \\
		1927.21 & 19.45 & 13.81$\pm$0.05 & ... & ... & ... & ... & ... \\
		1951.62 & 19.45 & ... & ... & ... & ... & 14.38$\pm$0.05 & ... \\
\hline
\hline
		\end{tabular}
		\\
		$a$. Taken from Maselli et al.\cite{Maselli2014}.
\end{center}
\end{table}

\newpage
\pagestyle{empty}
\begin{table}[h!]
\caption{ {\bf White band emission interpolated by {\it Swift}-UVOT narrow bands.} These data (except for the last column) have been plotted in the bottom panel of Extended data Fig. \ref{fig:GRB130427_SED}. \\}
\begin{center}
\label{tab:130427Awh}
	\begin{tabular}{cccccccccc}
		\hline
		\hline
		Epoch & T-T$_0$ & Exp & V & B & U & W1 & M2 & W2 & White$^a$ \\
    	 & (s) & (s) & (AB) & (AB) & (AB) & (AB) & (AB) & (AB) & (AB) \\
		\hline
		\hline
		1 & 515.57 & 19.44 & 12.32$\pm$0.04 & 11.51$\pm$0.35$^b$ & 12.31$\pm$0.30$^b$ & 12.82$\pm$0.04 & 12.89$\pm$0.04 & 13.05$\pm$0.04 & 12.62$\pm$0.44 \\
        2 & 688.49 & 19.45 & 12.59$\pm$0.04 & 12.79$\pm$0.04 & 12.93$\pm$0.06 & 13.18$\pm$0.04 & 13.20$\pm$0.04 & 13.45$\pm$0.04 & 12.95$\pm$0.06 \\
        3 & 860.19 & 19.45 & 12.82$\pm$0.04 & 13.00$\pm$0.04 & 13.11$\pm$0.04 & 13.41$\pm$0.04 & 13.44$\pm$0.04 & 13.75$\pm$0.04 & 13.19$\pm$0.09 \\
        4 & 1187.86 & 19.44 & 13.28$\pm$0.05 & 13.42$\pm$0.04 & 13.58$\pm$0.04 & 13.80$\pm$0.04 & 13.85$\pm$0.04 & 14.08$\pm$0.04 & 13.60$\pm$0.06 \\
        5 & 1359.98 & 19.45 & 13.48$\pm$0.05 & 13.63$\pm$0.04 & 13.77$\pm$0.04 & 13.98$\pm$0.04 & 14.02$\pm$0.05 & 14.18$\pm$0.04 & 13.77$\pm$0.05 \\
        6 & 1532.32 & 19.44 & 13.65$\pm$0.05 & 13.78$\pm$0.04 & 13.86$\pm$0.04 & 14.15$\pm$0.04 & 14.19$\pm$0.05 & 14.37$\pm$0.04 & 13.93$\pm$0.05 \\
        7 & 1704.19 & 19.44 & 13.75$\pm$0.05 & 13.88$\pm$0.04 & 14.02$\pm$0.04 & 14.28$\pm$0.04 & 14.27$\pm$0.05 & 14.41$\pm$0.04 & 14.03$\pm$0.05 \\
        8 & 1877.72 & 19.44 & 13.80$\pm$0.05 & 13.90$\pm$0.04$^c$ & 14.08$\pm$0.04$^c$ & 14.29$\pm$0.04$^c$ & 14.32$\pm$0.05 & 14.50$\pm$0.05 & 14.07$\pm$0.05 \\
\hline
\hline
		\end{tabular}
		\\
		$a$. Interpolated White band AB magnitude of GRB 130427A. To derive the intrinsic count rate in a 5 arcsec aperture, galactic extinction $A_{\rm WH}=0.0875$ have been accounted. Fitting uncertainties and standard deviation of fitting residuals contribute to uncertainties have been considered. \\
		$b$. At early phase, there is an additional radiation component, hence these 2 data points are excluded from SED fitting algorithm. \\
		$c$. These data points are results of extrapolation, hence they are excluded from SED fitting algorithm as well.
\end{center}
\end{table}

\newpage
\pagestyle{empty}
\begin{table}[h!]
	\caption{{\bf Photon count rates measured in aperture and halo ring methods in White band.} Sensitivity correction factors are 1.175 and 1.102 for GRB 220101A field and GRB 130427A field, respectively. The factor from count rate to flux is 0.01327 mJy/(count/s) for white band. These data points have been plotted in the top panel of  Extended data Fig. \ref{fig:apering}.}
\label{tab:ringwh}
\begin{center}

	\begin{tabular}{cccccccc}
	\hline
	\hline
	T-T$_{\rm 0}$ & Exposure & $\dot{N}_{\rm ring}^{\rm tot,raw}$ & $\dot{N}_{\rm ring}^{\rm bkg,raw}$ & COI$_{\rm tot(bkg)}$ & LSS & $\dot{N}_{\rm ring}$ & $\dot{N}_{\rm aper}$ \\
	(s) & (s) & (count/s) & (count/s) & & & (count/s) & (count/s) \\
	\hline
	\hline
	\multicolumn{8}{c}{GRB 220101A} \\
	\hline
	165.95 & 27.56 & 83.08$\pm$1.86 & 74.07$\pm$1.18 & 1.033(1.029) & 0.998 & 11.23$\pm$2.75 & 257.17$\pm$22.69 \\
	209.76 & 58.67 & 79.73$\pm$1.25 & 73.84$\pm$0.81 & 1.031(1.029) & 0.998 & 7.28$\pm$1.85 & 150.41$\pm$5.44 \\
	\hline
	\hline
	\multicolumn{8}{c}{GRB 130427A$^a$} \\
	\hline
	515.57 & 19.44 & 139.77$\pm$3.05 & 72.21$\pm$1.21 & 1.056(1.028) & 0.997 & 80.66$\pm$4.00 & 2030.98$\pm$831.79$^b$ \\
	688.49 & 19.45 & 133.49$\pm$2.98 & 72.09$\pm$1.21 & 1.054(1.028) & 0.996 & 73.20$\pm$3.90 & 1493.50$\pm$86.74 \\
	860.19 & 19.45 & 120.21$\pm$2.83 & 70.98$\pm$1.21 & 1.048(1.028) & 0.996 & 58.19$\pm$3.69 & 1204.35$\pm$101.10 \\
	1187.86 & 19.44 & 102.65$\pm$2.30 & 71.17$\pm$1.06 & 1.041(1.028) & 0.996 & 36.99$\pm$3.00 & 826.44$\pm$43.24 \\
	1359.98 & 19.45 & 92.88$\pm$2.19 & 71.45$\pm$1.06 & 1.037(1.028) & 0.996 & 25.09$\pm$2.86 & 703.53$\pm$29.41 \\
	1532.32 & 19.44 & 96.50$\pm$2.23 & 71.19$\pm$1.04 & 1.038(1.028) & 0.997 & 29.68$\pm$2.90 & 608.85$\pm$30.53 \\
	1704.19 & 19.44 & 92.26$\pm$2.18 & 71.03$\pm$1.05 & 1.037(1.028) & 0.997 & 24.86$\pm$2.84 & 553.88$\pm$26.34 \\
	1877.72 & 19.44 & 89.60$\pm$2.15 & 71.26$\pm$1.05 & 1.036(1.028) & 0.998 & 21.47$\pm$2.81 & 531.31$\pm$26.60 \\
	\hline
	\hline
	RA & DEC & $\dot{N}_{\rm ring}^{\rm tot,raw}$ & $\dot{N}_{\rm ring}^{\rm bkg,raw}$ & COI$_{\rm tot(bkg)}$ & LSS & $\dot{N}_{\rm ring}$ & $\dot{N}_{\rm aper}$ \\
	(J2000) & (J2000) & (count/s) & (count/s) & & & (count/s) & (count/s) \\
	\hline
	\hline
	\multicolumn{8}{c}{stars in GRB 220101A field$^c$.} \\
	\hline 
	00:05:43.983 & +31:47:20.11 & 80.83$\pm$0.76 & 74.00$\pm$0.49 & 1.032(1.029) & 1.006 & 8.58$\pm$1.14 & 168.01$\pm$4.10 \\
	00:05:33.844 & +31:42:10.45 & 81.85$\pm$0.75 & 74.00$\pm$0.48 & 1.032(1.029) & 1.014 & 9.94$\pm$1.12 & 208.39$\pm$6.07 \\
	00:05:26.211 & +31:48:43.76 & 83.42$\pm$1.06 & 73.99$\pm$0.67 & 1.033(1.029) & 0.996 & 11.73$\pm$1.56 & 230.06$\pm$7.61 \\
\hline
\hline
	\end{tabular}
	\\
	
	$a$. $\dot{N}_{\rm aper}$ is derived from SED. \\
	$b$. This data is not fitted since U-band exposures were saturated around this exposure, hence it could be unreliable(see Extended data Fig. \ref{fig:GRB130427_SED}). \\
	$c$. These data are measured with the first 150 second White band exposure in window timing mode.
\end{center}
\end{table}
	
\newpage
\pagestyle{empty}
\begin{table}[h!]
    \caption{{\bf Photon count rates measured in aperture and halo ring methods in v band.} The large scale structure correction factor and the sensitivity correction factor are 1.001 and 1.056, respectively. The factor from count rate to flux is 0.25491 mJy/(count/s) for V band. These data points used have been plotted in the bottom panel of Extended data Fig.  \ref{fig:apering}.} 
\label{tab:ringv}
\begin{center}
	\begin{tabular}{cccccccc}
		\hline
		\hline
		T-T$_0$ & Exp & $\dot{N}_{\rm ring}^{\rm tot,raw}$ & $\dot{N}_{\rm ring}^{\rm bkg,raw}$ & COI$_{\rm tot(bkg)}$ & $\dot{N}_{\rm ring}$ & ${{\rm Mag}_{\rm aper}}^a$ & ${\dot{N}_{\rm aper}}^b$ \\
    	(s) & (s) & (count/s) & (count/s) & & (count/s) & (AB) & (count/s)\\
		\hline
		\hline
		\multicolumn{8}{c}{080319B V band measurements with Wo\'{z}nika et al. as reference} \\
		\hline
		180.60 & 9.84 & 74.86$\pm$1.19 & 15.11$\pm$0.20 & 1.030(1.006) & 65.43$\pm$1.35 & 10.06$\pm$0.02 & 1349.02$\pm$24.85 \\
		193.52 & 9.84 & 65.04$\pm$1.39 & 14.95$\pm$0.20 & 1.026(1.006) & 54.63$\pm$1.56 & 10.23$\pm$0.02 & 1152.44$\pm$21.23 \\
		206.25 & 9.85 & 54.36$\pm$1.50 & 14.77$\pm$0.19 & 1.021(1.006) & 43.01$\pm$1.67 & 10.44$\pm$0.02 & 952.40$\pm$17.54 \\
		218.97 & 9.85 & 52.19$\pm$1.52 & 14.94$\pm$0.20 & 1.020(1.006) & 40.43$\pm$1.68 & 10.55$\pm$0.02 & 861.43$\pm$15.87 \\
		231.70 & 9.84 & 47.80$\pm$1.53 & 14.35$\pm$0.19 & 1.019(1.006) & 36.23$\pm$1.69 & 10.72$\pm$0.02 & 731.85$\pm$13.48 \\
		244.43 & 9.84 & 42.70$\pm$1.53 & 14.32$\pm$0.19 & 1.017(1.006) & 30.68$\pm$1.68 & 10.88$\pm$0.02 & 635.65$\pm$11.71 \\
		257.45 & 9.84 & 39.72$\pm$1.52 & 14.38$\pm$0.19 & 1.015(1.006) & 27.36$\pm$1.67 & 11.00$\pm$0.02 & 567.05$\pm$10.45 \\
		270.38 & 9.84 & 38.63$\pm$1.51 & 14.77$\pm$0.19 & 1.015(1.006) & 25.75$\pm$1.66 & 11.12$\pm$0.02 & 509.59$\pm$9.39 \\
		283.30 & 9.84 & 36.73$\pm$1.50 & 14.34$\pm$0.19 & 1.014(1.006) & 24.15$\pm$1.64 & 11.23$\pm$0.02 & 457.95$\pm$8.44 \\
		296.23 & 9.84 & 35.39$\pm$1.49 & 14.10$\pm$0.19 & 1.014(1.005) & 22.94$\pm$1.63 & 11.36$\pm$0.02 & 408.53$\pm$7.53 \\
		309.15 & 9.84 & 31.18$\pm$1.45 & 14.37$\pm$0.19 & 1.012(1.006) & 18.09$\pm$1.58 & 11.52$\pm$0.02 & 351.58$\pm$6.48 \\
		322.08 & 9.84 & 28.19$\pm$1.41 & 14.05$\pm$0.19 & 1.011(1.005) & 15.20$\pm$1.54 & 11.61$\pm$0.02 & 323.91$\pm$5.97 \\
		334.80 & 9.84 & 28.53$\pm$1.41 & 14.00$\pm$0.19 & 1.011(1.005) & 15.62$\pm$1.54 & 11.72$\pm$0.02 & 291.08$\pm$5.36 \\
		360.05 & 29.52 & 24.58$\pm$0.78 & 14.14$\pm$0.11 & 1.010(1.005) & 11.21$\pm$0.85 & 11.91$\pm$0.02 & 244.35$\pm$4.50 \\
		395.50 & 29.52 & 23.49$\pm$0.77 & 13.98$\pm$0.11 & 1.009(1.005) & 10.21$\pm$0.84 & 12.20$\pm$0.02 & 187.25$\pm$3.45 \\
		431.04 & 29.53 & 22.41$\pm$0.76 & 14.19$\pm$0.11 & 1.009(1.005) & 8.82$\pm$0.82 & 12.40$\pm$0.02 & 156.18$\pm$2.88 \\
		466.29 & 29.52 & 21.29$\pm$0.74 & 14.14$\pm$0.11 & 1.008(1.005) & 7.67$\pm$0.81 & 12.54$\pm$0.02 & 137.03$\pm$2.52 \\
		502.44 & 29.52 & 20.02$\pm$0.73 & 13.90$\pm$0.11 & 1.008(1.005) & 6.55$\pm$0.79 & 12.74$\pm$0.02 & 137.03$\pm$2.52 \\
		537.68 & 29.52 & 19.13$\pm$0.72 & 13.96$\pm$0.11 & 1.007(1.005) & 5.53$\pm$0.78 & 12.80$\pm$0.02 & 114.29$\pm$2.11 \\
		\hline
		\hline
		\multicolumn{8}{c}{080319B V band measurements with Page et al. as reference} \\
		\hline
		189.92 & 29.49 & 67.15$\pm$0.79 & 14.97$\pm$0.11 & 1.026(1.006) & 56.97$\pm$0.88 & 10.07$\pm$0.26 & 1335.42$\pm$319.79 \\
		224.89 & 39.36 & 49.06$\pm$0.76 & 14.58$\pm$0.10 & 1.019(1.006) & 37.38$\pm$0.84 & 10.44$\pm$0.29 & 949.77$\pm$253.68 \\
		269.89 & 49.21 & 38.26$\pm$0.67 & 14.48$\pm$0.09 & 1.015(1.006) & 25.67$\pm$0.74 & 10.89$\pm$0.38 & 627.51$\pm$219.62 \\
		322.39 & 54.13 & 29.98$\pm$0.61 & 14.06$\pm$0.08 & 1.012(1.005) & 17.13$\pm$0.67 & 11.60$\pm$0.74 & 326.30$\pm$222.40 \\
		\hline
		\hline
		\multicolumn{8}{c}{080319B V band measurements with \textit{HEASoft}} \\
		\hline
		322.39 & 54.13 & 29.98$\pm$0.61 & 14.06$\pm$0.08 & 1.012(1.005) & 17.13$\pm$0.67 & 11.75$\pm$0.02 & 284.20$\pm$5.24$^c$ \\
		357.39 & 14.77 & 23.47$\pm$1.09 & 14.15$\pm$0.16 & 1.009(1.005) & 10.00$\pm$1.18 & 11.91$\pm$0.04 & 245.26$\pm$9.04 \\
		372.39 & 14.76 & 24.23$\pm$1.10 & 14.37$\pm$0.16 & 1.009(1.006) & 10.58$\pm$1.20 & 12.10$\pm$0.04 & 205.88$\pm$7.58 \\
		387.39 & 14.77 & 24.79$\pm$1.11 & 14.16$\pm$0.16 & 1.010(1.005) & 11.40$\pm$1.21 & 12.14$\pm$0.04 & 198.44$\pm$7.31 \\
		402.40 & 14.77 & 22.70$\pm$1.08 & 13.84$\pm$0.15 & 1.009(1.005) & 9.51$\pm$1.17 & 12.21$\pm$0.04 & 186.05$\pm$6.85 \\
		417.40 & 14.76 & 22.44$\pm$1.07 & 14.46$\pm$0.16 & 1.009(1.006) & 8.56$\pm$1.17 & 12.35$\pm$0.04 & 163.54$\pm$6.02 \\
		432.40 & 14.77 & 22.51$\pm$1.07 & 14.14$\pm$0.16 & 1.009(1.005) & 8.97$\pm$1.17 & 12.44$\pm$0.04 & 150.53$\pm$5.55 \\
		447.40 & 14.76 & 20.96$\pm$1.05 & 13.89$\pm$0.15 & 1.008(1.005) & 7.58$\pm$1.14 & 12.49$\pm$0.04 & 143.75$\pm$5.30 \\
		462.40 & 14.77 & 21.64$\pm$1.06 & 14.21$\pm$0.16 & 1.008(1.005) & 7.96$\pm$1.15 & 12.51$\pm$0.04 & 141.13$\pm$5.20 \\
		477.40 & 14.76 & 21.09$\pm$1.05 & 14.22$\pm$0.16 & 1.008(1.005) & 7.36$\pm$1.14 & 12.60$\pm$0.04 & 129.90$\pm$4.79 \\
		492.39 & 14.77 & 21.44$\pm$1.06 & 13.95$\pm$0.15 & 1.008(1.005) & 8.03$\pm$1.15 & 12.62$\pm$0.04 & 127.53$\pm$4.70 \\
		507.39 & 14.76 & 19.91$\pm$1.03 & 13.98$\pm$0.15 & 1.008(1.005) & 6.35$\pm$1.12 & 12.77$\pm$0.04 & 111.08$\pm$4.09 \\
		522.39 & 14.77 & 19.06$\pm$1.01 & 13.82$\pm$0.15 & 1.007(1.005) & 5.60$\pm$1.10 & 12.79$\pm$0.04 & 109.05$\pm$4.02 \\
		537.40 & 14.77 & 19.94$\pm$1.03 & 13.76$\pm$0.15 & 1.008(1.005) & 6.62$\pm$1.12 & 12.88$\pm$0.04 & 100.37$\pm$3.70 \\
		552.40 & 14.76 & 19.08$\pm$1.01 & 14.00$\pm$0.16 & 1.007(1.005) & 5.43$\pm$1.10 & 12.89$\pm$0.04 & 99.45$\pm$3.66 \\
		567.30 & 14.57 & 18.62$\pm$1.01 & 14.19$\pm$0.16 & 1.007(1.005) & 4.75$\pm$1.10 & 12.94$\pm$0.04 & 94.98$\pm$3.50$^d$ \\
		719.60 & 19.47 & 16.60$\pm$0.84 & 13.88$\pm$0.13 & 1.006(1.005) & 2.90$\pm$0.91 & 13.35$\pm$0.04 & 65.11$\pm$2.40$^d$ \\
			1073.88 & 196.67 & 15.55$\pm$0.26 & 14.04$\pm$0.04 & 1.006(1.005) & 1.61$\pm$0.28 & 14.13$\pm$0.02 & 31.74$\pm$0.58$^d$ \\
		1273.77 & 196.77 & 15.34$\pm$0.25 & 14.16$\pm$0.04 & 1.006(1.005) & 1.26$\pm$0.28 & 14.52$\pm$0.02 & 22.16$\pm$0.41$^d$ \\
\hline
\hline
		\end{tabular}
		\\
		$a$. Magnitudes are taken from  Wo\'{z}nika et al.\cite{2009ApJ...691..495W} and Page et al.\cite{Page2013}.
		It is not necessary to take account for 
		the very small($\sim$ 0.01) difference between Vega magnitude and AB magnitude in V band.\\
		$b$. Only values in the last sub table 080319B V band measurements with \textit{HEASoft} are directly measured, others are all inferred values(i.e. $\dot{N}_{\rm int}$). \\
		$c$. This exposure is close to saturation, and Page et al.\cite{Page2013} derived a photometry result with readout streak method, which is consistent with the aperture photometry result given by \textit{HEASoft}. \\
		$d$. These points are not plotted in Fig 3 and not used in fitting algorithm as well.
\end{center}
\end{table}

\begin{table}
\caption{ {\bf Photometry for {\it Swift} UVOT observations of GRB 220101A.} } 
\label{tab:observation}
\begin{tabular}{lllllll}
\hline \hline
    Filter	& T$_{\rm start}$  & T$_{\rm end}$  & T$_{\rm exp}$	&Signal$^a$	&Sky & Mag$^b$ \\
	& second	 & second	 & second	& count/s	& count/s/pixel	& (AB) \\ 
\hline 
\hline

v	& 70.94	& 80.61	& 9.52	& $6.137\pm1.074$	&0.0313	& $16.12\pm0.19$ \\

white	&91.96	&93.62	&1.64	&$21.74\pm4.74$	&0.0145	&$17.75	\pm 0.24	$\\
white	&93.64	&97.62	&3.93	&$50.37\pm4.79$	&0.0148	&$16.83	\pm 0.10	$\\
white	&97.63	&101.62	&3.94	&$123.7\pm8.9$	&0.0150	&$15.86	\pm 0.08	$\\
white	&101.63	&105.63	&3.94	&$360.4\pm30.3$	&0.0147	&$14.70	\pm 0.09	$\\

white	&105.64	&109.62	&3.93	&$(634.4\pm171.8)^c$	&0.0146	&$(14.08	\pm 0.29)$	\\
white	&109.63	&113.63	&3.94	&$(765.0\pm175.6)^c$	&0.0147	&$(13.88	\pm 0.25)$	\\
white	&113.64	&117.62	&3.93	&$(1033\pm184.4)^c$	&0.0150	&$(13.56	\pm 0.19)$	\\
white	&117.63	&121.63	&3.94	&$(544.8\pm168.7)^c$	&0.0146	&$(14.25	\pm 0.34)$	\\
white	&121.64	&125.62	&3.93	&$(570.2\pm170.7)^c$	&0.0147	&$(14.20	\pm 0.33)$	\\
white	&125.63	&129.62	&3.94	&$(374.0\pm165.0)^c$	&0.0148	&$(14.66	\pm 0.48)$	\\
white	&129.64	&133.62	&3.93	&$(411.7\pm167.3)^c$	&0.0149	&$(14.55	\pm 0.44)$	\\
white	&133.63	&137.62	&3.94	&$(383.1\pm165.9)^c$	&0.0149	&$(14.63	\pm 0.47)$	\\
white	&137.63	&141.62	&3.93	&$(333.5\pm163.1)^c$	&0.0146	&$(14.78	\pm 0.53)$	\\
white	&141.63	&145.62	&3.94	&$(381.1\pm163.3)^c$	&0.0145	&$(14.64	\pm 0.47)$	\\
white	&145.63	&149.63	&3.94	&$(346.4\pm163.6)^c$	&0.0147	&$(14.74	\pm 0.51)$	\\

white	&149.64	&153.62	&3.93	&$330.2\pm26.4$	&0.0148	&$14.79	\pm 0.09	$\\
white	&153.63	&157.63	&3.94	&$328.3\pm26.1$	&0.0147	&$14.80	\pm 0.09	$\\
white	&157.64	&161.62	&3.93	&$303.1\pm23.3$	&0.0149	&$14.89	\pm 0.08	$\\
white	&161.63	&165.63	&3.94	&$291.6\pm22.0$	&0.0145	&$14.93	\pm 0.08	$\\
white	&165.64	&169.62	&3.93	&$257.8\pm18.7$	&0.0147	&$15.06	\pm 0.08	$\\
white	&169.63	&173.62	&3.94	&$224.0\pm15.8$	&0.0149	&$15.21	\pm 0.08	$\\
white	&173.63	&177.62	&3.93	&$215.8\pm15.2$	&0.0146	&$15.26	\pm 0.08	$\\
white	&177.63	&181.62	&3.94	&$202.9\pm14.2$	&0.0148	&$15.32	\pm 0.08	$\\
white	&181.63	&185.62	&3.93	&$208.6\pm14.6$	&0.0149	&$15.29	\pm 0.08	$\\
white	&185.63	&189.62	&3.94	&$186.0\pm12.9$	&0.0147	&$15.42	\pm 0.08	$\\
white	&189.63	&193.63	&3.94	&$192.9\pm13.4$	&0.0146	&$15.38	\pm 0.08	$\\
white	&193.64	&197.62	&3.93	&$179.6\pm12.5$	&0.0147	&$15.45	\pm 0.08	$\\
white	&197.63	&201.62	&3.94	&$164.6\pm11.5$	&0.0147	&$15.55	\pm 0.08	$\\
white	&201.64	&205.62	&3.93	&$187.3\pm13.1$	&0.0149	&$15.41	\pm 0.08	$\\
white	&205.63	&209.62	&3.94	&$146.1\pm10.3$	&0.0145	&$15.68	\pm 0.08	$\\
white	&209.64	&213.62	&3.93	&$158.1\pm11.1$	&0.0148	&$15.59	\pm 0.08	$\\
white	&213.63	&217.62	&3.94	&$147.3\pm10.4$	&0.0149	&$15.67	\pm 0.08	$\\
white	&217.63	&221.63	&3.94	&$134.1\pm9.6$	&0.0149	&$15.77	\pm 0.08	$\\
white	&221.64	&225.62	&3.93	&$135.8\pm9.7$	&0.0147	&$15.76	\pm 0.08	$\\
white	&225.63	&229.63	&3.94	&$117.3\pm8.5$	&0.0145	&$15.92	\pm 0.08	$\\
white	&229.64	&233.62	&3.93	&$118.3\pm8.6$	&0.0146	&$15.91	\pm 0.08	$\\
white	&233.63	&237.63	&3.94	&$114.9\pm8.4$	&0.0143	&$15.94	\pm 0.08	$\\
white	&237.64	&239.56	&1.90	&$102.2\pm10.9$	&0.0145	&$16.07	\pm 0.12	$\\

\hline
H$/$L$^d$	&	&	&	&	&	& \\
\hline
u	&3627.5	&3827.3	&196.6	&$0.0194\pm0.1266$	&0.0384	&$>20.74$ \\
b	&3832.6	&4032.4	&196.6	&$0.1181\pm0.1624$	&0.0635	&$>19.95$ \\
white	&4037.3	&4237.1	&196.6	&$0.4286\pm0.1266$	&0.1187	&$21.90\pm0.32$ \\
w2	&4242.9	&4442.6	&196.6	&$-0.3402\pm0.0613$	&0.0012	&$>21.54$ \\
v	&4447.6	&4647.4	&196.6	&$0.4469\pm0.0617$	&0.0151	&$18.66\pm0.15$ \\
m2	&4652.4	&4852.1	&196.6	&$0.0015\pm0.0168$	&0.0006	&$>21.25$ \\
w1	&4857.5	&5057.3	&196.6	&$0.0099\pm0.0308$	&0.0019	&$>21.33$ \\
u	&5062.2	&5201.4	&137.0	&$0.0659\pm0.0730$	&0.0080	&$>20.90$ \\
\hline
white	&10266	&11051	&765	&$0.3408\pm0.0502$	&0.0602	&$22.15\pm0.16$ \\
v	&21543	&22361	&798	&$0.1786\pm0.0280$	&0.0141	&$19.66\pm0.17$ \\
white	&27727	&28549	&802	&$0.1635\pm0.0396$	&0.0594	&$22.95\pm0.26$ \\
v	&39724	&40112	&378	&$0.1419\pm0.0340$	&0.0138	&$19.91\pm0.26$ \\
\hline 
white & 44845 & 46039 & 1137 & $0.1662\pm0.0492$ &0.0601	& $22.93\pm0.36$ \\
white & 50868 & 61521 & 1528 & $0.1369\pm0.0530$ &0.1073	& $23.13\pm0.42$ \\
white & 66833 & 85013 & 5485 & $0.0202\pm0.0378$ &0.1000	& $>23.33$ \\
\hline
\hline
\end{tabular}

$a$. Signal photon count rates have been corrected for coincidence losses\cite{Li2006,Poole2008} and long-term sensitivity correction.

$b$. Magnitudes are based on {\it Swift}/UVOT zeropoints\cite{Poole2008}, errors are adjusted by a binomial distribution\cite{Kuin2008}, limiting magnitudes are $3\sigma$ upper limits. These values have not been corrected for the Galactic extinctions of $E(B-V)=0.0483$\cite{Schlafly2011}. 

$c$. These data have been analyzed in ring apertures as introduced in Methods.

$d$. Images taken before are in high resolution, our photometry is in 5$''$ aperture, after are in low resolution, our photometry is in 3$''$ aperture.

\end{table}

\begin{table}
\begin{center}
\caption{\bf Properties of  extremely bright GRB flares at the peak.}
\begin{tabular}{llllllllllll}
\hline \hline
    GRB	& $z$	& Band	& M$_{\rm peak}$  & A$_{\rm V,MW}$	& $\beta$ & A$_{\lambda}$$^a$ & DM$^c$	& $M_{\rm abs}$	&Ref. \\
	& 	&  & AB	 & 	 & $f_{\nu}\propto \nu^{-\beta}$	& 	& 	& AB	&\\ 
\hline 
\hline
990123	& 1.600	& White to V	& M$_{\rm V}=8.86\pm0.02$	& 0.04	& 0.67	& 0.04	& 45.42	& -36.60	& \cite{Akerlof1999}\\
050904	& 6.295	& White to 9500\AA	& M$=12.13\pm0.24$$^b$	& 0.16	& 1.0	& 1.25	& 48.97	& -38.09	& \cite{Boer2006}\\
080319B	& 0.937	& White to V	& M$_{\rm V}=5.34\pm0.04$	& 0.03	& 0.5	& 0.03	& 43.99	& -38.68	& \cite{Racusin2008}\\
220101A	& 4.618	& White to V	& M$_{\rm wh}=13.56\pm0.19$	& 0.15	& 0.7	&4.78	& 48.19	& -39.41	& This work \\
\hline
\hline
\end{tabular}
\label{tab:absmag}
\end{center}

$a$. A$_{\lambda}$= A$_{\lambda,{\rm MW}}+$A$_{\lambda,{\rm host}}+$A$_{\lambda,{\rm IGM}}$ is derived from photometric SED fit. \\
$b$. Converted from $f_{\lambda=9500{\rm \AA}}=17\pm4\times10^{-15}$ erg cm$^{-2}$ s$^{-1}$ \AA$^{-1}$.\\
$c$. Absolute Magnitude at the peak $M_{\rm peak,abs}=M_{\rm peak}-DM-A_{\lambda}$.

\end{table}



\end{document}